\documentclass[10pt,journal,final]{IEEEtran}
\IEEEoverridecommandlockouts
\newtheorem{theorem}{Theorem}
\newtheorem{lemma}{Lemma}
\newtheorem{remark}{Remark}
\usepackage{amsfonts,amssymb}
\usepackage{ifpdf}
\usepackage{cite}
\usepackage{stfloats}
\usepackage{bm}
\usepackage{color,xcolor}
\usepackage{subfigure}
\ifCLASSINFOpdf
   \usepackage[pdftex]{graphicx}
   \graphicspath{{../pdf/}{../jpeg/}}
   \DeclareGraphicsExtensions{.pdf,.jpeg,.png}
\else
   \usepackage[dvips]{graphicx}

   \graphicspath{{../eps/}}

   \DeclareGraphicsExtensions{.eps}
\fi

\usepackage[cmex10]{amsmath}

\usepackage{algorithm}
\usepackage{algorithmic}
\ifCLASSOPTIONcompsoc
\else
  \usepackage[caption=false,font=footnotesize]{subfig}
\fi

\usepackage{makecell}
\begin{document}
\title{Throughput Maximization for UAV-aided  Backscatter Communication Networks}

\author{Meng~Hua,~\IEEEmembership{Student Member,~IEEE,}
Luxi~Yang,~\IEEEmembership{Senior Member,~IEEE,}
Chunguo~Li,~\IEEEmembership{Senior Member,~IEEE,}
Qingqing~ Wu,~\IEEEmembership{Member,~IEEE,}
        and~A. Lee Swindlehurst,~\IEEEmembership{Fellow,~IEEE}

\thanks{2019 Personal use is permitted, but republication/redistribution requires IEEE permission. Manuscript received April   27, 2019; revised July    29, and accepted November  7, 2019. This work was supported by National Natural Science Foundation of China under Grant  61971128,  Grant 61372101, and Grant 61671144, Scientific Research Foundation of Graduate School of Southeast University  under Grand  YBPY1859 and China Scholarship Council (CSC) Scholarship, National High Technology Project of China  under 2015AA01A703,  Cyrus Tang Foundation Endowed Young Scholar Program under SEU-CyrusTang-201801.    The associate editor coordinating the review of this paper and approving it for publication was Kamel Tourki. (\emph{Corresponding author: Luxi Yang}.)}
\thanks{M. Hua,  C. Li and L. Yang are with the School of Information Science and Engineering, Southeast University, Nanjing 210096, China (e-mail: \{mhua,  chunguoli, lxyang\}@seu.edu.cn).}
\thanks{Q. Wu is  with the Department of Electrical and Computer Engineering, National University of Singapore, Singapore.(e-mail: elewuqq@nus.edu.sg). }
\thanks{A. L. Swindlehurst is with the Center for Pervasive Communications and Computing, University of California at Irvine, Irvine, CA 92697 USA (e-mail: swindle@uci.edu).}
\thanks{Part of this work has been accepted by IEEE Global Communications Conference 2019 \cite{Hua2019throughput}. }.
}
\maketitle
\begin{abstract}
This paper investigates  unmanned aerial vehicle (UAV)-aided backscatter communication (BackCom) networks, where  the UAV is leveraged to help  the backscatter device (BD) forward signals to the receiver. Based on the presence or absence of a  direct link between BD and receiver, two protocols, namely transmit-backscatter (TB) protocol and transmit-backscatter-relay (TBR) protocol, are proposed to utilize the UAV to assist the BD. In particular, we formulate the system throughput  maximization problems for the two  protocols  by jointly optimizing the time allocation, reflection coefficient and UAV trajectory. Different static/dynamic  circuit power consumption models for the two  protocols  are  analyzed.  The resulting optimization problems are shown to be  non-convex, which are challenging to solve. We first consider the dynamic circuit power consumption model, and  decompose the original problems into three sub-problems, namely time allocation optimization with fixed UAV trajectory and reflection coefficient, reflection coefficient  optimization with fixed UAV trajectory and time allocation, and  UAV trajectory optimization with fixed reflection coefficient and time allocation. Then, an efficient iterative algorithm is proposed for both protocols by leveraging the block coordinate descent method and  successive convex approximation (SCA) techniques. In addition, for   the static   circuit power consumption model, we obtain the optimal time allocation  with a given reflection coefficient and UAV trajectory and the optimal  reflection coefficient  with low computational complexity by using the Lagrangian dual method.  Simulation results show that the proposed  protocols are able to achieve significant throughput gains over the compared benchmarks.
\end{abstract}


\section{Introduction}
In  wireless powered communication networks (WPCNs),  energy-constrained sensors powered by radio frequency (RF) signals  have been widely investigated due to their ability to provide reliable energy to Internet of Thing (IoT) devices\cite{xiethrouhput2018,wu2018Spectral,ju2014throughput,bi2016wireless,bi2015wireless,zeng2017communications,xu2015comp}. In WPCNs, the sensor nodes harvest the transmitted signal energy, and then transmit the data to a receiver by generating radio waves with the help of  active RF  components. However, active RF  components integrated into the sensor nodes  can still consume significant  energy for data forwarding. More energy-efficient devices, called  backscatter devices (BDs), have received significant attention in the past decade as a  promising technique for  IoT \cite{lyu2018optimal,kang2018riding,lyu2018relay,hoang2017ambient}.  Compared with  traditional  wireless powered devices that transmit signals  to the receiver by generating radio waves with the help of  active RF  components, the circuit power consumption of  a BD integrated with passive components is several orders of magnitude lower, and can significantly prolong  IoT lifetimes \cite{liu2013ambient}, \cite{lyu2017wireless}. A typical application of backscatter communication (BackCom) for IoT is in RF identification (RFID) scenarios. More specifically, the RF reader first transmits the RF signals to a  passive tag,   the tag harvests energy from the RF reader signal to power its circuit,  and then forwards the information bits carried on the received RF sinusoidal signal  back to the reader by  adjusting  its load  impedance to change the amplitude and phase of its backscattered signal\cite{Boyer2014Backscatter}, \cite{Boyer2013Space}.

In \cite{liu2013ambient}, a BackCom system design was presented that leverages  ambient RF sources such as TV station, cellular base stations (BS), WiFi access point, etc.  Although there are many benefits in  backscatter-based communication systems, one must address a number of technical challenges  ranging from the circuit design of the tags to designing the required transmission protocols. For example, the authors in \cite{parks2013wireless} studied two RF energy harvesting circuit prototypes, and the  sensor measurements from the  two prototypes  were tested from different locations in a real-world scenario. The results showed that the most sensitive RF harvesting sensor node can be operated at $200\rm m$ from a traditional  BS. In \cite{lyu2018relay}, a cooperative BackCom relaying system was studied, and the system throughput was maximized via optimal time allocation. In \cite{lyu2018optimal}, the same authors studied a BackCom-aided duty cycle protocol in which the BD either remained in a   sleep  or active state, and the  throughput maximization problem was formulated by jointly optimizing the sleep/active state and reflection coefficient. A  backscatter-aided cognitive wireless powered network was investigated in \cite{lyu2018throughput}, where a hybrid harvest-then-transmit protocol was proposed, and the optimal time allocation for energy harvesting and backscatter communication was derived. In \cite{wang2015relay}, the authors studied a relay-assisted secure BackCom network for maximizing the secrecy rate of the system, and a sub-optimal low-complexity  relay selection strategy was obtained based on the distance between the forward and reverse links. There has been considerable work on BackCom systems like that described above which assumes the existence of available RF power sources such as TV or BS transmission towers. However, in remote or underdeveloped areas, no such power sources may be available for providing RF energy to IoT BDs.

In this paper, we address some of the challenges necessary for communication among IoT devices.  A promising solution involves leveraging the unmanned aerial vehicles (UAVs) to act as RF power resources  to  assist  communication between the BD and receiver.  UAVs have already received significant attention both from academia and industry for various applications such as energy transmission,   data collection, hot spot offloading and wireless communications \cite{xu2018uav,zhan2018energy,wang2018power,zeng2016wireless,lyu2018uav,zeng2016throughput,Joint2019meng,wu2018Joint,jiang2012optimization,han2009optimization,zhan2011wireless,power2018meng,zeng2017energy,spectrum2018wang}.  The UAV's flexible mobility can be exploited to design a trajectory that increases network throughput. For example, the work in \cite{lyu2018uav} proposed to use a UAV as a mobile BS to serve cell-edge users and  offload data from traditional BS. Their numerical results showed that the common throughput was significantly improved by optimizing the UAV trajectory, bandwidth allocation and user partitioning. A similar problem was addressed in \cite{zeng2016throughput}, where the UAV acted as a relay to assist data transmission, and its trajectory and the source/relay power allocation were optimized.   The results showed that the UAV trajectory provided significant  gains in terms of system throughput. A UAV integrated into satellite-based  cognitive terrestrial network was considered in \cite{Joint2019meng}, assuming  the  UAV and BS cooperatively serve a terrestrial user by sharing the licensed satellite network spectrum. This work  showed that by carefully designing the UAV trajectory and BS/UAV power allocation, the throughput  of secondary networks can be significantly improved. A multi-UAV enabled system for serving multiple users was also shown in \cite{wu2018Joint} to improve throughput by carefully designing the UAV trajectories and power allocation.


In this paper, we study a UAV-aided  BackCom network, in which  the UAV is leveraged to assist data transmission from BD to receiver.  Our work is different from \cite{xie2018throughput}, which considered a UAV-enabled WPCN where the UAV is used to first charge the sensors in the downlink and then receive data in the uplink. In our work, we exploit a UAV to improve the communication connectivity between the BD and receiver in scenarios both with and without a direct connection between the two. For  the first case, the direct link between BD and receiver is assumed to be available and modeled as  a channel consisting of both distance-dependent path-loss and small-scale
Rayleigh fading. We propose a  transmit-backscatter (TB) protocol for this case where the UAV first transmits signals to the BD, and the BD directly reflects the signals to the receiver by adjusting the load impedance of BD to change the amplitude and phase of the backscattered signal.  For  the second case, the direct link between BD and receiver is unavailable caused by severe blockage. We propose a  transmit-backscatter-relay (TBR) protocol for this case, in which the UAV transmits signals to the BD, the BD  returns the signals back to the UAV, and the UAV then decodes the signals and forwards the signals to the receiver. Therefore, the resulting problems for maximizing the  capacity of  BackCom networks for these two cases are distinctly  different, which are discussed separately later in this paper.  Nevertheless, the joint design of UAV trajectory, BD's backscattering time and reflecting coefficient  can significantly improve the  BackCom networks capacity for both two cases. On the one hand, the UAV can adjust its location to establish the  stronger  UAV-BD link. As a result, more energy can be harvested at BD for backscattering its own data to the receiver. On the other hand, due to the  practical constraints such as final/initial UAV location and  flying  period, the UAV is not capable of hovering above the BD all the time. Therefore, the BD's backscattering time and reflecting coefficient should  be adaptively designed according to  the movement of the UAV trajectory.  For example, when the UAV is far way from the BD, the BD's backscattering time and reflecting power should be  reduced, whereas  when the UAV hovers above the BD, the  BD's backscattering time and reflecting power will be increased. It is worth pointing out that there generally exists a trade-off for the reflection coefficient between energy harvesting and backscattering  rate  \cite{lyu2018optimal}.  Note that  a stronger reflection coefficient means that the backscattering rate increases but less energy is harvested, and  reducing the harvested energy also reduces the time required for data forwarding.

Motivated by above, our goal is  to maximize the BackCom networks capacity for both two cases by jointly optimizing the UAV trajectory,  time allocation and reflection coefficient over a finite flight period of the UAV, subject to the  UAV mobility and practical  harvest-backscatter constraints. Due to the BD's backscattering time is highly related to the BD's circuit power consumption. Therefore, the precise modeling on the  circuit power consumption for the system design is of  paramount importance. Specifically, we  consider two circuit power consumption models, namely static and dynamic models, in this paper.  For the static  model, the BD's circuit power consumption is fixed regardless of the transmission rate.  For the dynamic  model, the BD's circuit power consumption is  a function of  the transmission rate  \cite{kang2018riding}. The main contributions are summarized as follows:
\begin{itemize}
 \item We propose two protocols for the direct link available case and direct link unavailable case in the   UAV-aided BackCom networks. Compared to the existing BackCom, this paper is first to exploit the UAV to improve the BackCom networks capacity via joint optimization of UAV trajectory, time allocation and reflection coefficient.

 \item We develop a three-layer iterative algorithm to solve the resulting non-convex  optimization problems for both models by using the block coordinate descent method and successive convex approximation (SCA) techniques. Specifically,  we decompose the formulated problem into three sub-problems: time allocation optimization with fixed UAV trajectory and reflection coefficient, reflection coefficient  optimization with fixed UAV trajectory and time allocation, and  UAV trajectory   optimization with fixed reflection coefficient and time allocation. Based on the  solutions to the three  sub-problems, a block coordinate descent method is proposed for alternately optimizing time allocation, reflection coefficient and UAV trajectory  to maximize the total system throughput.
 \item We consider both static and dynamic power consumption models for the BD. For the static circuit power consumption model, where the value of circuit power consumption is constant, we derive the optimal time allocation in closed form, and the optimal reflection coefficient using a low-complexity Lagrangian dual method for both protocols.
\end{itemize}
This paper is organized as follows. Section II introduces the system model and problem formulation. Section III considers the first case where a direct link between BD and receiver is available.  Section IV considers the second case where a direct link is unavailable. Numerical results are presented in Section V, and  the conclusions are given in Section VI.
\begin{figure}[!t]
\centerline{\includegraphics[width=2.5in]{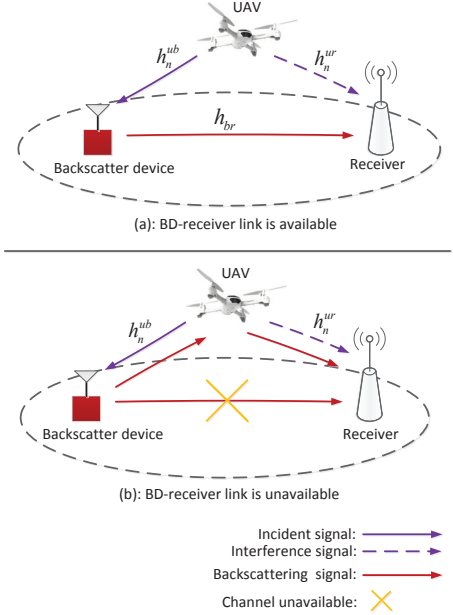}}
\caption{UAV-aided transmission in BackCom networks} \label{fig1}
\end{figure}
\section{System Model AND PROBLEM FORMULATION}
\subsection{System Model}
We consider a UAV-aided BackCom network that consists of one BD, one receiver and one UAV as shown in Fig.~\ref{fig1}. We assume that the UAV can freely adjust its heading to move with  a fixed altitude $H$. We assume that a finite flight period of the UAV is $T$. To make the problem tractable, the period $T$ is equally  divided into $N$\text{-}length time slots of  duration $\delta=T/N$. As a consequence, the horizontal location of UAV at time slot $n$ is  denoted as ${\bf q}_n$, $n\in {\cal N}=\{1,...,N\}$. The  horizontal coordinates of  the BD and receiver  are fixed at  ${\bf w}_b$ and ${\bf w}_r$, respectively.

The  UAV is likely to  establish  line-of-sight (LoS) links for both air-to-ground (A2G) and ground-to-air (G2A) channels as reported in \cite{matolak2015unmanned},\cite{khuwaja2018survey}. Therefore,  we model  the  A2G and G2A channels as  Rician fading \cite{zhan2018energy},\cite{you2019}.
Let $\sqrt {h_n^{ub}} $  and $\sqrt {h_n^{ur}} $ denote the UAV-BD and UAV-receiver channel coefficients at time slot $n$, respectively. The Rician fading consists of both distance-dependent path-loss and small-scale fading, which can be expressed as
\begin{align}
 \sqrt {h_n^{uf}} = \sqrt {{\theta _n^{uf}}}  {\tilde h}_n^{uf},
\end{align}
where $f\in\{b,r\}$, and ${\theta _{uf}}\left[ n \right]$ accounts for  large-scale channel attenuation  that depends on the path loss  and shadowing at time slot $n$, and  ${\tilde h}_n^{uf}$ is a complex-valued random variable with ${\mathbb E}\left[ {{{\| {\tilde h_n^{uf}}\|}^2}} \right]=1$ that represents  small-scale channel attenuation at time slot $n$. Specifically, ${{\theta _n^{uf}}}$ can be written as
\begin{align}
{\theta_n^{uf}}= \frac{{{\beta _0}}}{{{{\left\| {{{\bf{q}}_n} - {{\bf{w}}_f}} \right\|}^2} + {H^2}}},
\end{align}
where $\beta_0$ represents the reference channel gain at $d =1$ meter (m). Despite the result in \cite{ahmed2016importance} shown  that the  path loss exponents of A2G and G2A are  different, the gap between  A2G and G2A for  path loss exponent is very small. We thus assume that the two  path loss exponents are same with 2 that is consistent with the  most literatures adopted \cite{xu2018uav,zhan2018energy,lyu2018uav,zeng2016throughput,Joint2019meng,wu2018Joint,power2018meng,spectrum2018wang,you2019}. Although we assume that the  path-loss exponent is 2, it can be easily extended to other cases based on the results in Section II-A of  \cite{zeng2019accessing}.  Obviously, the value ${\theta _n^{uf}}$ depends on the distance between UAV and BD/receiver. The small-scale fading can be modeled as below
\begin{align}
\tilde h_n^{uf} = \sqrt {\frac{{{K_n}}}{{{K_n} + 1}}} {{\tilde h}} + \sqrt {\frac{1}{{{K_n} + 1}}} \tilde {\tilde h},
\end{align}
where ${{\tilde h}}$ denotes the deterministic LoS   channel coefficient with  $\| {{{\tilde h}}}\|=1$, and  $\tilde {\tilde h}$ is a circularly symmetric complex Gaussian random variable with mean zero and variance 1, and $K_n$ is a Rician factor at time slot $n$.
For a sufficiently  large value $K_n$, the channel model approximately equals to free-space path loss model. For $K_n=0$, the Rician channel model is simplified as Rayleigh channel model. The  Rician factor can be assumed to be invariant, namely $K=K_n$ for all $n$, for the following reasons. First, for the  rural areas and the limited serving range from several meters to dozens of meters in BackCom system, the Rician factor can be approximately treated to be  independent of the varying UAV locations. Second, for the long period time $T$, the most time for UAV is to stay stationary above the BD, thus  by assuming $K=K_n$ for all $n$ is reasonable.

We further assume that the channel model between BD and receiver follows Rayleigh fading with  channel power gain denoted by ${h_{br}} = {\beta _0}d_{br}^{ - m}\xi $, where $d_{br}$ is the distance between BD and receiver, $m$ denotes the path loss exponent, and $\xi$ is an exponentially distributed random variable with  mean 1.

In the following,  we introduce the main constraints that need to be considered for the Backcom networks.

\emph{Energy harvesting and circuit power consumption constraint:} Assuming that the UAV transmit power is $P$, the average received RF power by BD at time slot $n$ can be calculated as  $P\mathbb E\left[ {h_n^{ub}} \right]$. One part of the harvested energy is used to power  the BD's circuit power consumption, and the remaining harvested energy is used to backscatter its own data \cite{lyu2018optimal}, \cite{kang2018riding}. The harvested energy used for  powering  the BD's circuit is given by
\begin{align}
{E_n^b}\left[ n \right] = \eta \left( {1 - {a_n}} \right)P\mathbb E\left[ {h_n^{ub}} \right] = \eta \left( {1 - {a_n}} \right)P\theta _n^{ub},\label{Case1_energyharvesting}
\end{align}
where  $a_n$ denotes the reflection coefficient at time slot $n$ ($0\le a_n\le 1$), and $\eta$ represents the energy harvesting efficiency. Specially, $a_n=0$ indicates that all the harvested energy is used to power the  BD's circuit power consumption and  no power is left  to backscatter signals. $a_n=1$ indicates all the harvested energy is used to backscatter signals and  no power is left  to power the BD's circuit power consumption. Note that the reflection coefficient $a_n$ is controllable by  changing the impedance of an antenna in the presence of an incident signal \cite{lyu2018optimal},\cite{kang2018riding}.


In order for BD to work,   assuming that the signal processing delay at the BD is one time slot, we   have the following constraint
\begin{align}
\sum\limits_{i = 1}^n {{\varphi _{i{\rm{ + }}1}}P_{i{\rm{ + }}1}^e}  \le \sum\limits_{i = 1}^n {E_i^b\left[ n \right]} ,n{\rm{ = }}1, \ldots ,N - 1,\label{Case1_energyharvesting2}
\end{align}
where $\varphi_{i+1}$ denotes the portion of the BD's backscattering period for time slot $i+1$, and $P^e_{i+1}$ denotes the BD circuit power consumption at time slot $i+1$. The left hand side of \eqref{Case1_energyharvesting2} stands for the BD's circuit power consumption, and the right  hand side of \eqref{Case1_energyharvesting2} stands for the BD's harvested energy. The constraint \eqref{Case1_energyharvesting2} shows that  at each time slot $n$, the BD  can  be powered to work by  using the harvested energy in the  previous time slot. Note that as shown in  \cite{kang2018riding}, the BD's dynamic circuit energy consumption is given by
\begin{align}
P_{i{\rm{ + }}1}^e = {P_\varepsilon } + \mu {R_{i + 1}}, \label{Case1_dynamicenergymodel}
\end{align}
where $P_\varepsilon$ denotes the static energy consumption,  $\mu$ is a non-negative   weight factor \cite{kang2018riding}, and $R_{i+1}$ is the backscattering rate at time slot $i+1$. For $\mu=0$, namely static  circuit energy consumption model, the BD's  circuit energy consumption  is fixed with $P_\varepsilon$.
As a result, substituting   \eqref{Case1_energyharvesting} and \eqref{Case1_dynamicenergymodel} into \eqref{Case1_energyharvesting2}, we have
\begin{align}
\sum\limits_{i = 1}^n {{\varphi _{i + 1}}\left( {{P_\varepsilon } + \mu {R_{i + 1}}} \right)}  \le \sum\limits_{i = 1}^n {\eta \left( {1 - {a_i}} \right)} P\theta _i^{ub}.
\end{align}



\emph{UAV mobility constraint:} The UAV mobility is constrained  by its maximum flying speed, which implies that
\begin{align}
&\left\| {{\bf{q}}_{n+1} - {\bf{q}}_n} \right\| \le {V_{\max }}\delta,~n = 0, \ldots ,N-1,\notag\\
&{\bf{q}}_0 = {{\bf{q}}_{\rm I}},{\bf{q}}_N = {{\bf{q}}_{\rm{F}}},\label{Case1_constr2}
\end{align}
where $V_{\rm max}$ denotes the maximum UAV speed, ${{\bf{q}}_{\rm I}}$ and ${{\bf{q}}_{\rm{F}}}$ represent the UAV's initial and final location, respectively.

Next, we will separately discuss models of the  two cases, namely direct link available and unavailable case.
\subsubsection{Direct Link Available Between BD and Receiver}
In this case,  the UAV acts as a mobile energy transmitter to charge the BD and assist the BD for data backscatter transmission  as shown in Fig.~\ref{fig1}~(a). Different from the traditional devices that can generate the new RF radios, the BD can only leverage the ambient RF signal to backscatter its own data \cite{liu2013ambient}. Therefore, the BD cannot operate in the frequency division duplexing (FDD) model due to the fact that the BD uses the same frequency band for both the uplink and downlink.  Here, we use time-division duplexing (TDD) model, and we further assume that other co-channel RF sources are not present. We propose a transmit-backscatter (TB) protocol, which  consists of  two stages. In the first stage, the BD receives the broadcasting signals  from the UAV. In the second stage, the BD backscatters its own data by riding on the previously broadcasting signals to the receiver.

Specifically, in the first stage, at any time slot $n$, the received signal at the BD from the UAV is given by
\begin{align}
y_n^b = \sqrt P \sqrt {h_n^{ub}} {x_n}, \label{Case1_BD}
\end{align}
where $P$ is the UAV's maximum transmit power, and $x_n$ denotes  the UAV's transmitted signal at time slot $n$ with $\|x_n\|^2=1$. Note that the noise received at the BD is neglected because  the BD's circuit only includes passive components \cite{kang2018riding},\cite{wang2016ambient}, and \cite{qian2017noncoherent}.

In the second stage, at time slot $n+1$, the received signal at the receiver  from the BD is given by
\begin{align}
y_{n + 1}^r = \sqrt {h_{br}} \sqrt {{a_n}} y_n^b{c_n} + \sqrt P \sqrt {h_{n}^{ur}} {x_{n}} + {n_r}, \label{Case1_Receiver}
\end{align}
where $a_n$ denotes the BD's reflection coefficient at time slot $n$,  $n_r$  represents the received noise at the receiver with power $\sigma_r^2$,  and $c_n$ represents the BD's  own data with $\|c_n\|^2=1$. The term $\sqrt P \sqrt {h_{n}^{ur}} {x_{n}}$  denotes the received signal from the UAV at time slot $n$. Note that at the second stage, the UAV does not broadcast any signals since  the BD operates at half-duplex mode. Substituting \eqref{Case1_BD} into \eqref{Case1_Receiver}, we have
\begin{align}
y_{n + 1}^r{\rm{ = }}\sqrt {h_{br}} \sqrt {{a_n}} \sqrt P \sqrt {h_n^{ub}} {x_n}{c_n} + \sqrt P \sqrt {h_{n}^{ur}} {x_{n}} + {n_r}. \label{Case1_receiverV2}
\end{align}
The strength of the backscattering signal received at receiver  from the BD is generally  much lower than that received from the UAV. Thus, the successive
interference cancellation (SIC) technique can be applied. Specifically,  the receiver  first decodes the UAV signals  and then subtracts it from the combined signals before decoding its own signals \cite{lyu2018optimal,kang2018riding,lyu2018relay}. Thus, the  signal-plus-noise ratio (SNR) at the receiver at time slot $n+1$ can be expressed as
\begin{align}
\gamma _{n + 1}^r = \frac{{P{a_n}h_{br}h_n^{ub}}}{{\sigma _r^2}}. \label{Case1_expression_1}
\end{align}
Note that since the channel power gains ${h_{br}}$ and $h_n^{ub}$ are the random variables, the instantaneously  achievable rate is also a random variable, we thus pay attention to obtaining the expected communication throughput.  The expected  rate of the receiver   at time slot $n+1$ is given by
\begin{align}
R_{n + 1}^r = {\mathbb E}\left[ {{{\log }_2}\left( {1 + \gamma _{n + 1}^r} \right)} \right]{\rm{ }}. \label{Case1_expression_2}
\end{align}
The closed-form expressions of  $R_{n + 1}^r$  in \eqref{Case1_expression_2} and even its lower bound result of   $R_{n + 1}^r$   are  unsolvable  due to the
difficulty of deriving its probability distribution. To address this issue, one feasible approach is to use an approximation result of $R_{n + 1}^r$. We have the following Theorem:
\begin{theorem} \label{theorem1}
The approximation result of $R_{n + 1}^r$, denoted as $\hat R_{n + 1}^r$, is given by
\begin{align}
\hat R_{n + 1}^{r} = {\log _2}\left( {1 + \frac{{{W_{br}}{a_n}}}{{{{\left\| {{{\bf{q}}_n} - {{\bf{w}}_b}} \right\|}^2} + {H^2}}}} \right), \label{Case1_Ratelower}
\end{align}
where ${W_{br}} = \frac{{{e^{ - {\kappa _0}}}P{\beta _0}}}{{{\lambda _{br}}\sigma _r^2}}$.
\end{theorem}
\begin{IEEEproof}
Please refer to Appendix~\ref{appendix1}.
\end{IEEEproof}
The accuracy for the approximation in Theorem 1 will be evaluated in Section V under different parameters.

\subsubsection{Direct Link Unavailable Between BD and Receiver}
In this case, we consider the  scenario where a direct link between BD and receiver is unavailable.  The UAV acts as a relay  to assist the BD for backscattering  data to the receiver  using a decode-and-forward (DF) manner  as shown in Fig.~\ref{fig1} (b).  We propose a transmit-backscatter-relay (TBR) protocol, which consists of three stages. In the first stage, the UAV transmits broadcasting signals to the BD. In the second stage, the BD backscatters the own data  via riding on the previously broadcasting signals to the UAV. Note that in this stage, the UAV only receives the signals  and does not transmit broadcasting signals since the UAV operates in the half-duplex mode. In the third stage, the UAV acts as a relay to transmit the previously received BD's data to the receiver.
Specifically, in the first stage,  at time slot $n$, the UAV transmits signals to the BD, and  the received signal at BD  is given by
\begin{align}
z_n^b = \sqrt P \sqrt {h_n^{ub}} {x_n}, \label{Case2_BD}
\end{align}
where $P$ is the UAV's maximum transmit power, $x_n$ is the UAV broadcasting signal at time slot $n$ with $\|x_n\|^2=1$. In the second stage, at time slot $n+1$, the BD forwards its own data by riding on the previously UAV broadcasting signal to the UAV,  the received signal at the UAV  is given by
\begin{align}
z_{n + 1}^u = \sqrt {h_{n + 1}^{ub}} \sqrt {{a_n}} z_n^b{c_n} + {n_u},\label{Case2_UAV}
\end{align}
where $a_n$, $c_n$, and $n_u$  represent reflection coefficient, BD's transmit data, and received noise  at time slot $n$, respectively. The noise power of $n_u$ is  $\sigma_u^2$. Substituting \eqref{Case2_BD} into \eqref{Case2_UAV}, we have
\begin{align}
z_{n + 1}^u{\rm{ = }}\sqrt {P{a_n}h_{n + 1}^{ub}h_n^{ub}} {x_n}{c_n} + {n_u}.
\end{align}
Similar as previous discussion on  the direct link case, we are interested in the expected rate. The expected rate from BD to UAV at time slot $n+1$ is given by
\begin{align}
\check R_{n + 1}^u = \mathbb E\left[ {{{\log }_2}\left( {1 + \frac{{P{a_n}h_{n + 1}^{ub}h_n^{ub}}}{{\sigma _u^2}}} \right)} \right].\label{Case2_RU}
\end{align}
We assume that the UAV is capable of correctly decoding  the BD's signal $c_n$ at any  time slot $n+1$.  In the third stage, at time slot $n+2$, the UAV transmits its decoded BD's data to the receiver, the   received signal at the receiver  is given by
\begin{align}
z_{n + 1}^r{\rm{ = }}\sqrt {{P}} \sqrt {h_{n + 2}^{ur}} {s_{n+2}} + {\rm{ }}\sqrt {{P}} \sqrt {h_n^{ur}} {x_n} + {n_r},\label{Case2_receiver}
\end{align}
where $s_{n+2}$ denotes UAV's transmitted signal at time slot $n$, namely $s_{n+2}=c_n$. The term $\sqrt P \sqrt {h_n^{ur}} {x_n}$  denotes the received signal from the UAV at previous time slot $n$.  Similarly,  the SIC technique is also applied at the receiver, which has been   previously discussed in \eqref{Case1_receiverV2}.  The expected transmission rate from UAV to receiver  at time slot $n+2$ is thus  given by
\begin{align}
\check R_{n + 2}^r = \mathbb E\left[ {{{\log }_2}\left( {1 + \frac{{Ph_{n + 2}^{ur}}}{{\sigma _r^2}}} \right)} \right].\label{Case2_Rr_simulation}
\end{align}
To make the problem more tractable, we write $ {\check R}^u_{n+1}$ in \eqref{Case2_RU} as $\check R_{n + 1}^u = \mathbb E\left[ {{{\log }_2}\left( {1 + \frac{{P{a_n}{{\left( {h_{n + 1}^{ub}} \right)}^2}}}{{\sigma _u^2}}} \right)} \right]$   by assuming that $h^{ub}_{n+1} \cong h^{ub}_n$. Generally, it is  challenging to obtain the  probability distributions  of $\check R_{n + 1}^u$ and  $\check R_{n + 2}^r$. Similar as in Theorem \ref{theorem1}, the approximation results for  $\check R_{n + 1}^u$ and  $\check R_{n + 2}^r$, denoted as $R_{n + 1}^u$ and  $R_{n + 2}^r$, can be respectively given by
\begin{align}
R_{n + 1}^u = {\log _2}\left( {1 + \frac{{P{a_n}{{\left( {\theta _{n + 1}^{ub}} \right)}^2}}}{{\sigma _u^2}}} \right),\label{Case2_RU_approxi}
\end{align}
and
\begin{align}
R_{n + 2}^r = {\log _2}\left( {1 + \frac{{P\theta _{n + 2}^{ur}}}{{\sigma _r^2}}} \right).\label{Case2_Rr_approxi}
\end{align}
The derivations  for \eqref{Case2_RU_approxi} and \eqref{Case2_Rr_approxi} are  similar to that of   Theorem~\ref{theorem1}, and are omitted here for brevity. Section V  shows that the  approximation results match well with the numerical results.
\subsection{Problem Formulation}
For the  direct link available case, our objective is to maximize the sum of  ergodic capacity by jointly optimizing the UAV trajectory, time allocation and reflection coefficient under the  energy harvesting and circuit power consumption  constraint  as well as  UAV mobility constraint. Define ${\cal N}_1=\{1,2,..,N/2\}$. The problem is   formulated as follows.
\begin{align}
&\left( {{\rm{P1}}} \right)\mathop {\max }\limits_{a_{2n-1},\varphi_{2n},{\bf q}_n} \sum\limits_{n = 1}^{N/2} {\varphi_{2n}\hat R^{r}_ {2n }}\notag\\
&{\rm s.t.}~ \sum\limits_{i = 1}^n {{\varphi _{2i}}\left( {{P_\varepsilon } + \mu \hat R_{2i}^r} \right)}  \le \sum\limits_{i = 1}^n {\eta \left( {1 - {a_{2i - 1}}} \right)} P\theta _{2i - 1}^{ub},n \in {{\cal N}_1},\label{Case1_constr1}\\
&\quad 0 \le a_{2n-1}  \le 1,n\in {\cal N}_1,\label{Case1_constr3}\\
&\quad 0 \le \varphi_{2n } \le 1,n\in {\cal N}_1,\label{Case1_constr4}\\
&\quad\eqref{Case1_constr2},\notag
\end{align}
where \eqref{Case1_constr1} denotes the energy harvesting and circuit power consumption constraint  discussed in Section II.A, \eqref{Case1_constr3} and \eqref{Case1_constr4} represent the BD's backscattering  coefficient  and backscattering  time constraints, respectively.

Similarly, for the  direct link unavailable case, we aim  to maximize  the sum of ergodic capacity by jointly optimizing the UAV trajectory, time allocation and reflection coefficient under some specified constraints. Define ${\cal N}_2=\{1,2,..,N/3\}$. Mathematically, the optimization problem is formulated as
\begin{align}
&\left( {{\rm{P2}}} \right)\mathop {\max }\limits_{{a_{3n - 2}},{\varphi _{3n - 1}},{\bf{q}}\left[ n \right]} \sum\limits_{n = 1}^{{N \mathord{\left/
 {\vphantom {N 3}} \right.
 \kern-\nulldelimiterspace} 3}} {{\varphi _{3n - 1}}R_{3n - 1}^u}  \notag\\
&{\rm s.t.}~\sum\limits_{i = 1}^n {{\varphi _{3i - 1}}\left( {{P_\varepsilon } + \mu  R_{3i - 1}^r} \right)}  \le \notag\\
&\qquad\qquad\qquad\sum\limits_{i = 1}^n {\eta \left( {1 - {a_{3i - 2}}} \right)} P\theta _{3i - 2}^{ub},n \in {{\cal N}_2}, \label{Case2_const1_1}\\
&\qquad\sum\limits_{i = 1}^n {\varphi _{3i - 1}R_{3i - 1}^u}  \le \sum\limits_{i = 1}^n {R_{3i}^r},n \in {\cal N}_2,\label{Case2_information}\\
&\qquad 0 \le a_ {3n - 2}  \le 1,n \in {\cal N}_2,\label{Case2_const2}\\
&\qquad 0 \le \varphi_ {3n - 1} \le 1,n \in {\cal N}_2,\label{Case2_const3}\\
&\qquad \eqref{Case1_constr2}, \notag
\end{align}
where \eqref{Case2_const1_1} represents the energy harvesting and circuit power consumption constraint  discussed in Section II.A, and \eqref{Case2_information} denotes the information-casuality constraint by assuming that  the signal processing delay at the UAV is one time slot. \eqref{Case2_const2} and \eqref{Case2_const3} stand for the backscattering  coefficient and  backscattering  time constraints, respectively.

Problems $(\rm P1)$ and $(\rm P2)$ are  highly non-convex optimization problems where the optimization variables are intricately coupled in the objective function and constraints.  Specifically, first, the UAV trajectory, BD's backscattering time, and BD's reflecting coefficient are closely coupled in the objective function in $(\rm P1)$ or $(\rm P2)$, which  results in  non-convexity of $(\rm P1)$ or $(\rm P2)$. In addition, the constraints \eqref{Case1_constr1} and \eqref{Case2_const1_1} are also non-convex with respect to (w.r.t.)  UAV trajectory, BD's backscattering time as well as  BD's reflecting coefficient.  In generally, there is no standard method for solving such non-convex optimization problems optimally.  In Sections III and IV, we propose an alternating optimization algorithms for solving problems $(\rm P1)$ and $(\rm P2)$, respectively.

\section{Proposed Algorithm  For  Problem $(\rm P1)$} \label{section3}
In this section, we consider problem $(\rm P1)$ for the direct link available case. Problem $(\rm P1)$ is challenging to solve due to the non-convex objective function and constraint \eqref{Case1_constr1}. To this end, we decompose $(\rm P1)$ into three sub-problems, namely time allocation optimization with fixed UAV trajectory and reflection coefficient, reflection coefficient  optimization with fixed UAV trajectory and time allocation, and  UAV trajectory   optimization with fixed reflection coefficient and time allocation. Based on the  solutions to the three  sub-problems, a block coordinate descent method is proposed for alternately optimizing the  time allocation, reflection coefficient and UAV trajectory  to maximize the total system throughput.
\subsubsection{Time allocation optimization}
For any given UAV trajectory $\{{\bf q}_n\}$ and reflection coefficient $\{a_{2n-1}\}$, the time allocation $\{\varphi_{2n}\}$ of problem $(\rm P1)$ can be  optimized by solving the following problem
\begin{align}
&\left( {{\rm{P1}}{\rm{.1}}} \right)\mathop {\max }\limits_{\varphi _ {2n } } \sum\limits_{n = 1}^{N/2} {\varphi_{2n }{{\log }_2}\left( {1 + \frac{{{W_{br}}a_{2n-1} }}{{{{\left\| {{\bf{q}}_{ 2n-1} - {{\bf{w}}_b}} \right\|}^2} + {H^2}}}} \right)} \notag\\
&\qquad{\rm s.t.}~\eqref{Case1_constr1}, \eqref{Case1_constr4}.\notag
\end{align}
Since problem $(\rm P1.1)$ is a standard linear programming problem, it can be efficiently solved by interior point method \cite{boyd2004convex}.

\begin{theorem} \label{theorem2}
For  the static circuit energy consumption model, namely $\mu=0$,  if $\hat R^{r}_{2n}$ is a decreasing function with  $n \in{\cal N}_1$,  the optimal solution $\{\varphi_{2n}^*\}$ to problem $(\rm P1.1)$ is given by
\begin{align}
\varphi _{2n}^* = \left[ {\frac{{\eta \left( {1 - {a_{2n - 1}}} \right)P\theta_{2n - 1}^{ub}}}{{{P_c}}}} \right]_0^1, n\in {\cal N}_1. \label{Case_1timeallocation}
\end{align}
\end{theorem}
\begin{IEEEproof}
Please refer to Appendix~\ref{appendix2}.
\end{IEEEproof}
Theorem~\ref{theorem2} shows that the equality in \eqref{Case1_constr1} must hold at any time $n$, which means that the energy harvested by the BD  at time slot $n$ from the UAV will be thoroughly depleted for the BD's data backscattering transmission at time slot $n+1$. This transmission approach is similar to the time switching-based relaying (TSR) protocol of \cite{nasir2013relaying}.

\subsubsection{Reflection coefficient optimization}
For any given UAV trajectory $\{{\bf q}_n\}$ and  time allocation $\{\varphi_{2n}\}$,  the reflection coefficient $\{a_{2n-1}\}$ of problem $(\rm P1)$ can be optimized by solving the following problem
\begin{align}
&\left( {{\rm{P1}}.2} \right)\mathop {\max }\limits_{{a_{2n - 1}}} \sum\limits_{n = 1}^{N/2} {{\varphi _{2n}}{{\log }_2}\left( {1 + \frac{{{W_{br}}{a_{2n - 1}}}}{{{{\left\| {{{\bf{q}}_{2n - 1}} - {{\bf{w}}_b}} \right\|}^2} + {H^2}}}} \right)}  \notag\\
&\qquad{\rm s.t.}~\eqref{Case1_constr1}, \eqref{Case1_constr3}.\notag
\end{align}
Despite  the objective function ${{{\log }_2}\left( {1 + \frac{{{W_{br}}{a_{2n - 1}}}}{{{{\left\| {{{\bf{q}}_{2n - 1}} - {{\bf{w}}_b}} \right\|}^2} + {H^2}}}} \right)}$ is concave w.r.t. $a_{2n-1}$,  the constraint  \eqref{Case1_constr1} is non-convex w.r.t. $a_{2n-1}$. In general, there is no efficient method to obtain an optimal solution. In the following, we obtain an efficiently approximate solution to $(\rm P1.2)$ based on SCA techniques. It is observed that the term $\hat R_{2i}^{r}$ in the  left hand side (LHS) of  \eqref{Case1_constr1} is concave w.r.t. $a_{2n-1}$. To proceed, define $a^l_{2n-1}$ as the given time reflection coefficient at the $l$\text{-}th iteration, we have
\begin{align}
&{\hat R^{r}_{2n}} \le {\log _2}\left( {1 + \frac{{{W_{br}}a_{2n - 1}^l}}{{{{\left\| {{{\bf{q}}_{2n - 1}} - {{\bf{w}}_b}} \right\|}^2} + {H^2}}}} \right)+\notag\\
&\qquad\qquad A_{2n - 1}^l\left( {{a_{2n - 1}} - a_{2n - 1}^l} \right)\overset{\triangle}{ =} {\psi _{up}}\left( {\hat R_{2n}^{^{r}}} \right),\label{Case1_apprximation_timeallocation}
\end{align}
where $A_{2n - 1}^l = \frac{1}{{\ln 2}}\frac{{{W_{br}}}}{{{{\left\| {{{\bf{q}}_{2n - 1}} - {{\bf{w}}_b}} \right\|}^2} + {H^2} + {W_{br}}a_{2n - 1}^l}}$. Thus,  the  constraint \eqref{Case1_constr1}  can be  replaced as
\begin{align}
&\sum\limits_{i = 1}^n {{\varphi _{2i}}\left( {{P_\varepsilon } + \mu {\psi _{up}}\left( {\hat R_{2i}^{^{r}}} \right)} \right)}  \le \notag\\
&\qquad\qquad\qquad\qquad\sum\limits_{i = 1}^n {\eta \left( {1 - {a_{2i - 1}}} \right)} P\theta_{2i - 1}^{ub}, n \in {{\cal N}_1},\label{Case1_constr1NEWtimeallocation}
\end{align}
which is convex. As a  result, for  any feasible point $\{a^l_{2n-1}\}$, define the following optimization problem
\begin{align}
&\left( {{\rm{P1}}.3} \right)\mathop {\max }\limits_{{a_{2n - 1}}} \sum\limits_{n = 1}^{N/2} {{\varphi _{2n}}{{\log }_2}\left( {1 + \frac{{{W_{br}}{a_{2n - 1}}}}{{{{\left\| {{{\bf{q}}_{2n - 1}} - {{\bf{w}}_b}} \right\|}^2} + {H^2}}}} \right)}  \notag\\
&\qquad{\rm s.t.}~\eqref{Case1_constr3},\eqref{Case1_constr1NEWtimeallocation}.\notag
\end{align}
Based on the previous discussions, $(\rm P1.3)$ is a convex optimization problem that can be efficiently solved by standard convex optimization solvers \cite{grant2008cvx}. Then, $(\rm P1.2)$  can be approximately solved by successively updating the time allocation based on the optimal solution to $(\rm P1.3)$. In addition, it readily follows that the objective  of $(\rm P1.3)$ gives a lower bound to that of $(\rm P1.2)$.

For  the static circuit energy consumption model, namely $\mu=0$,  problem $(\rm P1.2)$ becomes a  convex optimization problem. It can easily verified that $(\rm P1.2)$ satisfies the slater's condition, as a result, its optimal solution can be obtained via solving the dual problem with low computational complexity \cite{boyd2004convex}.
\begin{lemma} \label{lemma1}
With the given dual variables ${\nu _n}>0$, $n=1,...,N/2$, corresponding to \eqref{Case1_constr1} with  $\mu=0$, the optimal reflection coefficient $\{a^*_{2n-1}\}$ for  $(\rm P1.2)$ is given by
\begin{align}
&a_{2n - 1}^* =\notag\\
&{\rm{ }}\left[ {\frac{{{\varphi _{2n}}}}{{\ln 2\eta P\theta_{2n - 1}^{ub}\sum\limits_{i = n}^{N/2} {{\nu _i}} }} - \frac{{{{\left\| {{{\bf{q}}_{2n - 1}} - {{\bf{w}}_b}} \right\|}^2} + {H^2}}}{{{W_{br}}}}} \right]_0^1.\label{Case_1statictimeallocation}
\end{align}
\end{lemma}
\begin{IEEEproof}
Please refer to Appendix~\ref{appendix3}.
\end{IEEEproof}
The dual problem  of $(\rm P1.2)$, denoted as $(\rm P1.2\text{-}D)$,  is defined as $\mathop {\min }\limits_{{v_n}} g\left( {{v_n}} \right)$.  This dual  problem can be solved by applying the subgradient method, which is guaranteed to  converge to a   globally optimal solution \cite{yu2006dual}.  The update rule for the dual variables $\{\nu_n\}$ is given by \cite{wu2015resource}
\begin{align}
&v_n^{t + 1} = \notag\\
&{\rm{ }}{\left[ {v_n^t - \pi \left( {\sum\limits_{i = 1}^n {\eta \left( {1 - {a_{2i - 1}}} \right)} P\theta_{2i - 1}^{ub} - \sum\limits_{i = 1}^n {{\varphi _{2i}}{P_c}} } \right)} \right]^ + },
\end{align}
where the  superscript $t$ denotes the iteration index, and $\pi$ is the positive step size.  In addition, the total computational complexity of using the Lagrange dual method is ${\cal O}{({K_\nu }\frac{N}{2})^2}$, where $\frac{N}{2}$ is number of dual variables, and $K_{\nu}$ represents the  number of iterations required for updating $\nu_n$ \cite{wu2015resource}.
\subsubsection{UAV trajectory  optimization}
For any given   reflection coefficient $\{a_{2n-1}\}$ and  time allocation $\{\varphi_{2n}\}$,   the  UAV trajectory $\{{\bf q}_n\}$ of problem $(\rm P1)$ can be  optimized by solving the following problem
\begin{align}
&\left( {{\rm{P1}}.4} \right)\mathop {\max }\limits_{{{\bf{q}}_n}} \sum\limits_{n = 1}^{N/2} {{\varphi _{2n}}{{\log }_2}\left( {1 + \frac{{{W_{br}}{a_{2n - 1}}}}{{{{\left\| {{{\bf{q}}_{2n - 1}} - {{\bf{w}}_b}} \right\|}^2} + {H^2}}}} \right)} \notag\\
&\qquad{\rm s.t.}~\eqref{Case1_constr2},\eqref{Case1_constr1}.\notag
\end{align}
Problem $(\rm P1.4)$ is a  non-convex optimization problem due to the non-convex objective function and constraint \eqref{Case1_constr1}. To tackle the non-convex objective function, the SCA technique is again applied. It can be observed that  ${{{\log }_2}\left( {1 + \frac{{{W_{br}}{a_{2n - 1}}}}{{{{\left\| {{{\bf{q}}_{2n - 1}} - {{\bf{w}}_b}} \right\|}^2} + {H^2}}}} \right)}$ is convex w.r.t. ${{{\left\| {{{\bf{q}}_{2n - 1}} - {{\bf{w}}_b}} \right\|}^2}}$, but it is not convex w.r.t. ${\bf q}_{2n-1}$. Taking  the first-order Taylor expansion at  any feasible point ${{{\left\| {{{\bf{q}}^l_{2n - 1}} - {{\bf{w}}_b}} \right\|}^2}}$ for the $l\text{-}$th iteration, we have the following inequality
\begin{align}
&\hat R_{2n}^{^{r}} \ge {\log _2}\left( {1 + \frac{{{W_{br}}{a_{2n - 1}}}}{{{{\left\| {{\bf{q}}_{_{2n - 1}}^l - {{\bf{w}}_b}} \right\|}^2} + {H^2}}}} \right) - B_{2n - 1}^l \times \notag\\
&\left( {{{\left\| {{{\bf{q}}_{2n - 1}} - {{\bf{w}}_b}} \right\|}^2} - {{\left\| {{\bf{q}}_{_{2n - 1}}^l - {{\bf{w}}_b}} \right\|}^2}} \right) \overset{\triangle}{=} {\Psi _{lb}}\left( {\hat R_{2n}^{^{r}}} \right),
\end{align}
where $B_{2n - 1}^l$ is given  in \eqref{Case1_P1_4_B} (see the top of the next page).
\newcounter{mytempeqncnt0}
\begin{figure*}
\normalsize
\setcounter{mytempeqncnt0}{\value{equation}}
\begin{align}
B_{2n - 1}^l = \frac{1}{{\ln 2}}\frac{{{W_{br}}{a_{2n - 1}}}}{{\left( {{{\left\| {{\bf{q}}_{_{2n - 1}}^l - {{\bf{w}}_b}} \right\|}^2} + {H^2}} \right)\left( {{W_{br}}{a_{2n - 1}} + {{\left\| {{\bf{q}}_{_{2n - 1}}^l - {{\bf{w}}_b}} \right\|}^2} + {H^2}} \right)}}.\label{Case1_P1_4_B}
\end{align}
\hrulefill 
\vspace*{4pt} 
\end{figure*}
To handle the non-convex  constraint \eqref{Case1_constr1}, we first reformulate it by introducing slack variables $\{{ y}_{2n-1}\}$ as
\begin{align}
&\sum\limits_{i = 1}^n {{\varphi _{2i}}\left( {{P_\varepsilon } + \mu {{\log }_2}\left( {1 + \frac{{{W_{br}}{a_{2n - 1}}}}{{{y_{2n - 1}} + {H^2}}}} \right)} \right)} \notag\\
&\qquad\quad\quad\quad\qquad\le \sum\limits_{i = 1}^n {\eta \left( {1 - {a_{2i - 1}}} \right)} P\theta_{2i - 1}^{ub},n \in {{\cal N}_1}, \label{Case1_constr1trajectory1}
\end{align}
with the additional constraint
\begin{align}
{\left\| {{{\bf{q}}_{2n - 1}} - {{\bf{w}}_b}} \right\|^2} \ge {y_{2n - 1}},n \in {{\cal N}_1}.\label{Case1_constr1trajectory2}
\end{align}
Note that both constraints \eqref{Case1_constr1trajectory1} and \eqref{Case1_constr1trajectory2} are non-convex. Similarly, to handle the non-convexity of \eqref{Case1_constr1trajectory1},  we  have the following inequality for \eqref{Case1_constr1trajectory1}  by applying the first-order Taylor expansion at the given point ${{{\left\| {{\bf{q}}_{_{2n - 1}}^l - {{\bf{w}}_b}} \right\|}^2}}$  in the $l$\text{-}th iteration,
\begin{align}
&\sum\limits_{i = 1}^n {{\varphi _{2i}}\left( {{P_\varepsilon } + \mu {{\log }_2}\left( {1 + \frac{{{W_{br}}{a_{2n - 1}}}}{{{y_{2n - 1}} + {H^2}}}} \right)} \right)} \notag\\
&\qquad\qquad\quad\quad\le \sum\limits_{i = 1}^n {\eta \left( {1 - {a_{2i - 1}}} \right)} P\theta_{2i - 1}^{ub,lb},n \in {{\cal N}_1},  \label{Case1_constr1trajectory1NEW}
\end{align}
where $\theta_{2n - 1}^{ub,lb} = \frac{{{\beta _0}}}{{{{\left\| {{\bf{q}}_{_{2n - 1}}^l - {{\bf{w}}_b}} \right\|}^2} + {H^2}}} - \frac{{{\beta _0}}}{{{{\left( {{{\left\| {{\bf{q}}_{_{2n - 1}}^l - {{\bf{w}}_b}} \right\|}^2} + {H^2}} \right)}^2}}}\left( {{{\left\| {{{\bf{q}}_{2n - 1}} - {{\bf{w}}_b}} \right\|}^2} - {{\left\| {{\bf{q}}_{_{2n - 1}}^l - {{\bf{w}}_b}} \right\|}^2}} \right)$.
To handle the non-convexity of \eqref{Case1_constr1trajectory2} w.r.t. ${\bf q}_{2n-1}$,  we  can also obtain the following inequality for  \eqref{Case1_constr1trajectory2} by applying the first-order Taylor expansion at the given point ${\bf q}^l_{2n-1}$ in the $l$\text{-}th iteration,
\begin{align}
&{\left\| {{\bf{q}}_{_{2n - 1}}^l - {{\bf{w}}_b}} \right\|^2} + 2{\left( {{\bf{q}}_{_{2n - 1}}^l - {{\bf{w}}_b}} \right)^T} \times \left( {{{\bf{q}}_{2n - 1}} - {\bf{q}}_{_{2n - 1}}^l} \right) \notag\\
&\qquad\qquad\qquad\qquad\qquad\qquad\qquad\ge {y_{2n - 1}},n \in {{\cal N}_1}. \label{Case1_constr1trajectory2NEW}
\end{align}
As a  result, for  any feasible  points $\left\{ {{{\left\| {{\bf{q}}_{2n - 1}^l - {{\bf{w}}_b}} \right\|}^2}} \right\}$ and $\{{\bf q}^l_{2n-1}\}$, problem  $(\rm P1.4)$ is approximated as
\begin{align}
&\left( {{\rm{P1}}{\rm{.5}}} \right)\mathop {\max }\limits_{{\bf{q}}_n,y_{2n-1}} \sum\limits_{n = 1}^{N/2} {\varphi_ {2n}{\Psi _{lb}}\left( {\hat R^{r}_ {2n}} \right)} \notag\\
&\qquad{\rm s.t.} ~\eqref{Case1_constr2},~\eqref{Case1_constr1trajectory1NEW},~\eqref{Case1_constr1trajectory2NEW}.\notag
\end{align}
It can be readily verified that the objective function and all  the constraints  are convex, and thus $(\rm P1.5)$  can be solved by standard convex optimization solvers \cite{grant2008cvx}. Then, problem $(\rm P1.4)$  can be approximately solved by successively updating the UAV trajectory  based on the optimal solution to problem $(\rm P1.5)$. In addition, it readily follows that the objective of $(\rm P1.5)$ provides a lower bound to that of problem $(\rm P1.4)$.
\subsubsection{Overall algorithm}
Based on the solutions to its three sub-problems above, we alternately optimize the three sub-problems in an iterative way, and  a locally optimal solution to problem $(\rm P1)$ can be obtained. The details of the alternating  optimization algorithm are summarized in Algorithm~\ref{alg1}.
\begin{algorithm}[H]\label{alg1}
\caption{Alternating optimization algorithm}
\label{alg1}
\begin{algorithmic}[1]
\STATE  \textbf{Initialize} UAV  trajectory  ${{\bf{q}}_{_{2n - 1}}^l}$,   reflection coefficient ${a_{2n - 1}^l}$, and set $l \leftarrow 0$  as well as tolerance $\epsilon > 0$ .
\STATE  \textbf{repeat}.
\STATE  \quad Solve problem  $(\rm P1.1)$ for  given  $\left\{ {{\bf{q}}_{_{2n - 1}}^l,a_{2n - 1}^l} \right\}$, and \\
\quad denote the optimal solution as   $\{\varphi^{l+1}_{2n}\}$.
\STATE  \quad Solve problem  $(\rm P1.2)$ for  given   $\left\{ {{\bf{q}}_{_{2n - 1}}^l,\varphi_{2n }^{l+1}} \right\}$, and \\
\quad  denote the optimal solution as   $\{a^{l+1}_{2n-1}\}$.
\STATE  \quad Solve problem  $(\rm P1.4)$ for  given  $\{\varphi_{2n}^{l+1} ,a_{2n - 1}^{l+1} \}$, and \\
\quad    denote the optimal solution as  ${{\bf{q}}_{_{2n - 1}}^{l+1}}$. \\
\STATE \quad $l \leftarrow l + 1$.
\STATE \textbf{until} the fractional increase of the objective value of $\left( {{\rm{ P1}}} \right)$ is less than tolerance  $\epsilon$.
\end{algorithmic}
\end{algorithm}
The convergence of Algorithm~\ref{alg1} is shown as follows: Define $R\left( {\varphi _{2n}^l,a_{2n-1}^l,{\bf{q}}_n^l} \right)$ as the objective value of $(\rm P1)$  in the $l$\text{-}th iteration,  $R_{ref}^{lb}\left( {\varphi _{2n}^l,a_{2n-1}^l,{\bf{q}}_n^l} \right)$ as the objective value of $(\rm P1.3)$  in the $l$\text{-}th iteration, and  $R_{trj}^{lb}\left( {\varphi _{2n}^l,a_{2n-1}^l,{\bf{q}}_n^l} \right)$ as the objective value of $(\rm P1.5)$  in the $l$\text{-}th iteration. In the $l\text{+}1$\text{-}th iteration, in   step 3 of Algorithm~\ref{alg1}, we have
\begin{align}
R\left( {\varphi _{2n}^l,a_{2n - 1}^l,{\bf{q}}_n^l} \right)\overset{a} {\le} R\left( {\varphi _{2n}^{l + 1},a_{2n - 1}^l,{\bf{q}}_n^l} \right). \label{convengece_1}
\end{align}
 The inequality (a) holds since ${\varphi _{2n}^{l + 1}}$ is the optimal solution to problem $(\rm P1.1)$. In  step 4, it follows that
 \begin{align}
 R\left( {\varphi _{2n}^{l + 1},a_{2n - 1}^l,{\bf{q}}_n^l} \right) &\overset{b}{=} R_{ref}^{lb}\left( {\varphi _{2n}^{l + 1},a_{2n - 1}^l,{\bf{q}}_n^l} \right)\notag\\
 &\overset{c}{\le} R_{ref}^{lb}\left( {\varphi _{2n}^{l + 1},a_{2n - 1}^{l + 1},{\bf{q}}_n^l} \right)\notag\\
 &\overset{d}{\le} R\left( {\varphi _{2n}^{l + 1},a_{2n - 1}^{l + 1},{\bf{q}}_n^l} \right),\label{convengece_2}
 \end{align}
where equality (b) holds since the first-order Taylor expansion at point ${a_{2n - 1}^l}$ is tight in \eqref{Case1_apprximation_timeallocation}, and  inequality (c) holds since  ${a_{2n - 1}^{l + 1}}$ is the optimal solution to problem $(\rm P1.3)$, and inequality (d) holds since the objective value of  $(\rm P1.3)$ is lower bounded by $(\rm P1.2)$ at any given point ${a_{2n - 1}^{l + 1}}$. In step 5, we have
\begin{align}
R\left( {\varphi _{2n}^{l + 1},a_{2n - 1}^{l + 1},{\bf{q}}_n^l} \right) &= R_{trj}^{lb}\left( {\varphi _{2n}^{l + 1},a_{2n - 1}^{l + 1},{\bf{q}}_n^l} \right)\notag\\
 &\le R_{trj}^{lb}\left( {\varphi _{2n}^{l + 1},a_{2n - 1}^{l + 1},{\bf{q}}_n^{l + 1}} \right)\notag\\
  &\le R\left( {\varphi _{2n}^{l + 1},a_{2n - 1}^{l + 1},{\bf{q}}_n^{l + 1}} \right),\label{convengece_3}
\end{align}
which is similar to  \eqref{convengece_2}. Based on \eqref{convengece_1}\text{-}\eqref{convengece_3}, we have
\begin{align}
R\left( {\varphi _{2n}^l,a_{2n - 1}^l,{\bf{q}}_n^l} \right) \le R\left( {\varphi _{2n}^{l + 1},a_{2n - 1}^{l + 1},{\bf{q}}_n^{l + 1}} \right),
\end{align}
which shows that the  objective value of $(\rm P1)$ is non-decreasing over the iterations in Algorithm~\ref{alg1}. In addition, the  objective value of $(\rm P1)$ is  upper-bound by a finite value due to the limited flight time. As such, Algorithm~\ref{alg1} is guaranteed to converge.

Next, we analyze  the complexity of Algorithm~\ref{alg1}. In step 3 of  Algorithm~\ref{alg1}, the sub-problem $(\rm P1.1)$ is a linear optimization problem, and can be solved by interior point method with computational complexity ${\cal O}\left( {\sqrt {\frac{N}{2}} \frac{1}{\varepsilon }} \right)$, where $\frac{N}{2}$ denotes the number of  variables, and $\varepsilon$ represents the iterative accuracy \cite{gondzio1996computational}. In step 4 of  Algorithm~\ref{alg1}, since problem $(\rm P1.2)$  involves logarithmic form, the  complexity for solving $(\rm P1.2)$ is ${\cal O}\left( {{L_1}{{\left( {\frac{N}{2}} \right)}^{3.5}}} \right)$, where $L_1$ is the number of  iterations required to update reflection coefficient \cite{zhang2019securing}. In step 5 of  Algorithm 1,  problem $(\rm P1.5)$ is a convex quadratic programming problem, which involves $\frac{5N}{2}$ scalar real decision variables, and thus the computational complexity of $(\rm P1.4)$ is ${\cal O}\left( {{L_2}{{\left( {\frac{{5N}}{2}} \right)}^3}} \right)$ \cite{spectrum2018wang}. Therefore, the overall computational complexity of Algorithm~\ref{alg1} is ${\cal O}\left( {{L_3}\left( {\sqrt {\frac{N}{2}} \frac{1}{\varepsilon } + {L_1}{{\left( {\frac{N}{2}} \right)}^{3.5}} + {L_2}{{\left( {\frac{{5N}}{2}} \right)}^3}} \right)} \right)$ with $L_3$ being the number of iterations of Algorithm~\ref{alg1}. This result shows that the complexity of Algorithm~\ref{alg1} is polynomial in the worst scenario.

\section{Proposed  Algorithm  For  Problem $(\rm P2)$}
In this section, we consider problem $(\rm P2)$ for the direct link unavailable case. To address  $(\rm P2)$, the following remark is used.
\begin{remark} \label{remark1}
The inequality  constraint \eqref{Case2_information} always holds. At any time, the uplink transmission rate $R^u_{3n-1}$ is  a  two-hop  transmission, while the downlink transmission rate $R^r_{3n}$ is  a one-hop  transmission, and thus $R^u_{3n-1} \ll R^r_{3n}$ $\left({\left( {\theta _{3n - 1}^{ub}} \right)^2} \ll \theta _{3n}^{ur}\right)$. This indicates that the information-causality constraint is always satisfied, and we can omit it in our formulated problem. This result has also been verified in Section V.
\end{remark}

Based on Remark~\ref{remark1}, problem $(\rm P2)$ can be simplified as follows
\begin{align}
&\left( {{\rm{P3}}} \right)\mathop {\max }\limits_{{a_{3n - 2}},{\varphi _{3n - 1}},{\bf{q}}\left[ n \right]} \sum\limits_{n = 1}^{{N \mathord{\left/
 {\vphantom {N 3}} \right.
 \kern-\nulldelimiterspace} 3}} {{\varphi _{3n - 1}}R_{3n - 1}^u}  \notag\\
 &{\rm s.t.}~\eqref{Case1_constr2}, \eqref{Case2_const1_1}, \eqref{Case2_const2},\eqref{Case2_const3}.\notag
\end{align}
Problem $(\rm P3)$ is still challenging to solve  due to the non-convex  objective function and constraint \eqref{Case2_const1_1}. As before discussed for $(\rm P1)$,  we decompose $(\rm P3)$ into three sub-problems: time allocation optimization with fixed UAV trajectory and reflection coefficient, reflection coefficient  optimization with fixed UAV trajectory and time allocation, and  UAV trajectory   optimization with fixed reflection coefficient and time allocation. Then, a block coordinate descent method is still used to maximize the sum of  system throughput by alternately optimizing time allocation, reflection coefficient and UAV trajectory.
\setcounter{subsubsection}{0}
\subsubsection{Time allocation optimization}
For any given UAV trajectory $\{{\bf q}_n\}$ and reflection coefficient $\{a_{3n-2}\}$, the time allocation $\{\varphi_{3n-1}\}$ can be optimized by solving the following problem
\begin{align}
&\left( {{\rm{P3}}{\rm{.1}}} \right)\mathop {\max }\limits_{{\varphi _{3n - 1}}} \sum\limits_{n = 1}^{{N \mathord{\left/
 {\vphantom {N 3}} \right.
 \kern-\nulldelimiterspace} 3}} {{\varphi _{3n - 1}}R_{3n - 1}^u}   \notag\\
&{\rm s.t.}~\eqref{Case2_const1_1}, \eqref{Case2_const3}.\notag
\end{align}
Problem $(\rm P3.1)$ is a linear programming problem, which can  be efficiently solved by the interior point method\cite{boyd2004convex}.
\begin{theorem} \label{theorem4}
For  the static circuit energy consumption model, namely $\mu=0$, if $R^{u}_{3n-1}$ is a decreasing function with  $n \in {\cal N}_2$,  the optimal solution $\{\varphi_{3n-1}^*\}$ to problem $(\rm P3.1)$ is
\begin{align}
\varphi _{3n - 1}^* = \left[ {\frac{{\eta \left( {1 - {a_{3n - 2}}} \right)P\theta_{3n - 2}^{ub}}}{{{P_c}}}} \right]_0^1, n\in{\cal N}_2.
\end{align}
\end{theorem}
\begin{IEEEproof}
The proof is similar to that of Theorem ~\ref{theorem2} in Appendix~\ref{appendix2}.
\end{IEEEproof}
Similar to that of  Theorem~\ref{theorem2},  Theorem~\ref{theorem4}  also shows the same result that  the energy harvested by the BD at time slot $n$ from the UAV will be thoroughly depleted for the BD's backscattering  at time slot $n+1$.
\subsubsection{Reflection coefficient optimization}
For any given UAV trajectory $\{{\bf q}_n\}$ and  time allocation $\{\varphi_{3n-1}\}$,  the reflection coefficient $\{a_{3n-2}\}$ can be optimized by solving the following problem
\begin{align}
&\left( {{\rm{P3}}{\rm{.2}}} \right)\mathop {\max }\limits_{{a_{3n - 2}}} \sum\limits_{n = 1}^{{N \mathord{\left/
 {\vphantom {N 3}} \right.
 \kern-\nulldelimiterspace} 3}} {{\varphi _{3n - 1}}R_{3n - 1}^u}   \notag\\
&{\rm s.t.}~\eqref{Case2_const1_1}, \eqref{Case2_const2}.\notag
\end{align}
Problem $(\rm P3.2)$ is non-convex due to the constraint  \eqref{Case2_const1_1}. However, we observe  that the left hand side of \eqref{Case2_const1_1} is concave w.r.t. $a_{3n-2}$.  By employing a Taylor expansion at  any feasible point $a^l_{3n-2}$ at the $l$\text{-}th iteration, a convex upper bound ${\tilde R}^u_{3n-1}$ for $R^u_{3n-1}$ can be expressed as
\begin{align}
&\tilde R_{3n - 1}^u = {\log _2}\left( {1 + \frac{{a_{3n - 2}^lP{{\left( {\theta_{3n - 1}^{ub}} \right)}^2}}}{{\sigma _u^2}}} \right)\notag\\
&\qquad\qquad\qquad\qquad\qquad+ \tilde A_{3n - 1}^l\left( {{a_{3n - 2}} - a_{3n - 2}^l} \right), \label{Case2_RUNEW}
\end{align}
where $\tilde A_{3n - 1}^l = \frac{1}{{\ln 2}}\frac{{P{{\left( {\theta_{3n - 1}^{ub}} \right)}^2}}}{{a_{3n - 2}^lP{{\left( {\theta_{3n - 1}^{ub}} \right)}^2} + \sigma _u^2}}$. With \eqref{Case2_RUNEW}, the constraint \eqref{Case2_const1_1} can be transformed as
\begin{align}
&\sum\limits_{i = 1}^n {{\varphi _{3i - 1}}\left( {{P_\varepsilon } + \mu \tilde R_{3i - 1}^u} \right)}  \le \notag\\
&\qquad \qquad \qquad\qquad  \sum\limits_{i = 1}^n {\eta \left( {1 - {a_{3i - 2}}} \right)} P\theta_{3i - 2}^{ub},n \in {{\cal N}_2}.\label{Case2_const1NEW}
\end{align}
As a result, for any given point $a^l_{3n-2}$, we have
\begin{align}
&\left( {{\rm{P3}}{\rm{.3}}} \right)\mathop {\max }\limits_{{a_{3n - 2}}} \sum\limits_{n = 1}^{{N \mathord{\left/
 {\vphantom {N 3}} \right.
 \kern-\nulldelimiterspace} 3}} {{\varphi _{3n - 1}}R_{3n - 1}^u}  \notag\\
&{\rm s.t.}~\eqref{Case2_const2}, \eqref{Case2_const1NEW}.\notag
\end{align}
Problem $(\rm P3.3)$ is a convex optimization problem that can be efficiently solved by standard methods \cite{grant2008cvx}. Then, problem $(\rm P3.2)$  can be approximately solved by successively updating the time allocation based on the optimal solution to problem $(\rm P3.3)$.

For  the static circuit energy consumption model, namely $\mu=0$,  problem $(\rm P3.2)$ becomes a  convex optimization problem.  The globally optimal reflection coefficient to $(\rm P3.2)$  can be obtained  with low-complexity by using Lagrangian dual method, which is discussed below.
\begin{lemma} \label{lemma2}
With the given dual variables ${\bar\nu _n}>0$, $n=1,...,N/3$,  corresponding to   \eqref{Case2_const1_1}, the optimal reflection coefficient $\{a^*_{3n-2}\}$ for  $(\rm P3.2)$ is given by
\begin{align}
&a_{3n - 2}^* =\notag\\
&\left[ {\frac{{{\varphi _{3n - 1}}}}{{\ln 2\eta P\theta_{3n - 2}^{ub}\sum\limits_{i = n}^{N/3} {{{\bar \nu }_i}} }} - \frac{{\sigma _u^2}}{{P{{\left( {\theta_{3n - 1}^{ub}} \right)}^2}}}} \right]_0^1, n\in {\cal N}_2.
\end{align}
\end{lemma}
\begin{IEEEproof}
The proof is similar to that of Lemma ~\ref{lemma1} in Appendix~\ref{appendix3}.
\end{IEEEproof}
The dual problem  of $(\rm P3.2)$, denoted as $(\rm P3.2\text{-}D)$, can be obtained  by applying the subgradient method \cite{yu2006dual}.  The update rule for the dual variables $\{{\bar \nu}_n\}$ is given by
\begin{align}
&{\bar \nu}_n^{t + 1} = \notag\\
&{\rm{ }}{\left[ {{\bar \nu}_n^t - \pi \left( {\sum\limits_{i = 1}^n {\eta \left( {1 - {a_{3i - 2}}} \right)} P\theta_{3i - 2}^{ub} - \sum\limits_{i = 1}^n {{\varphi _{3i - 1}}{P_c}} } \right)} \right]^ + },
\end{align}
where the  superscript $t$ denotes the iteration index, and $\pi$ is the positive step size.  In addition, the total computational complexity of using Lagrange dual method is ${\cal O}{({K_{\bar \nu }}\frac{N}{3})^2}$, where $K_{\bar \nu}$ represents the number of iterations required for updating ${\bar \nu}_n$ \cite{wu2015resource}.

\subsubsection{UAV trajectory optimization}
For any given   reflection coefficient $\{a_{3n-2}\}$ and  time allocation $\{\varphi_{3n-1}\}$,   the  UAV trajectory $\{{\bf q}_n\}$ of problem $(\rm P3)$ can be  optimized by solving the following problem
\begin{align}
&\left( {{\rm{P3}}{\rm{.4}}} \right)\mathop {\max }\limits_{{\bf{q}}\left[ n \right]} \sum\limits_{n = 1}^{{N \mathord{\left/
 {\vphantom {N 3}} \right.
 \kern-\nulldelimiterspace} 3}} {{\varphi _{3n - 1}}R_{3n - 1}^u}  \notag\\
&{\rm s.t.}~\eqref{Case1_constr2}, \eqref{Case2_const1_1}.\notag
\end{align}
Note that the objective function  and constraint \eqref{Case2_const1_1} of problem $(\rm P3.4)$  are non-convex. In general, there is no efficient method
to obtain the optimal solution. In the following, we solve it by using SCA techniques. By introducing slack variables $\{t_{3n-1}\}$ and $\{s_{3n-1}\}$,  problem $(\rm P3.4)$ can be equivalently formulated as
\begin{align}
&\left( {{\rm{P3}}.{\rm{5}}} \right)\mathop {\max }\limits_{{t_{3n - 1}},{s_{3n - 1}},{{\bf{q}}_n}} \sum\limits_{n = 1}^{{N \mathord{\left/
 {\vphantom {N 3}} \right.
 \kern-\nulldelimiterspace} 3}} {{\varphi _{3n - 1}}{{\log }_2}\left( {1 + \frac{{P{a_{3n - 2}}\beta _0^2}}{{\sigma _u^2{{\left( {{t_{3n - 1}}} \right)}^2}}}} \right)}   \notag\\
 &{\rm s.t.}~\sum\limits_{i = 1}^n {{\varphi _{3i - 1}}\left( {{P_\varepsilon } + \mu {{\log }_2}\left( {1 + \frac{{P{a_{3i - 2}}\beta _0^2}}{{\sigma _u^2{{\left( {{s_{3i - 1}}} \right)}^2}}}} \right)} \right)} \notag\\
 & \qquad \qquad \qquad \le \sum\limits_{i = 1}^n {\eta \left( {1 - {a_{3i - 2}}} \right)} P\theta_{3i - 2}^{ub},n \in {{\cal N}_2}, \label{Case2_P3_5_const1}\\
 &\qquad {\left\| {{{\bf{q}}_{3n - 1}} - {{\bf{w}}_b}} \right\|^2} + {H^2} \le {t_{3n - 1}},n\in {\cal N}_2,\label{Case2_P3_5_const2}\\
 &\qquad {\left\| {{{\bf{q}}_{3n - 1}} - {{\bf{w}}_b}} \right\|^2} + {H^2} \ge {s_{3n - 1}},n\in {\cal N}_2,\label{Case2_P3_5_const3}\\
&\qquad \eqref{Case1_constr2}.\notag
\end{align}
It can be verified that at the optimal solution to problem $(\rm P3.5)$, the constraints \eqref{Case2_P3_5_const2} and \eqref{Case2_P3_5_const3} are met with equality, since otherwise we can always decrease $t_{3n-1}$ and increase $s_{3n-1}$ to obtain a larger objective. In the objective function of problem $(\rm P3.5)$, since ${{{\log }_2}\left( {1 + \frac{{P{a_{3n - 2}}\beta _0^2}}{{\sigma _u^2{{\left( {{t_{3n - 1}}} \right)}^2}}}} \right)}$ is convex w.r.t. $t_{3n-1}$, we have the  inequality in \eqref{Case2_P3_5_ObjectNEW} by taking the first-order Taylor expansion at  any feasible  point $t^l_{3n-1}$, where ${C_{3n - 1}} = \frac{{P{a_{3n - 2}}\beta _0^2}}{{\sigma _u^2}}$ (see the top of the next page).  Note that  in  constraint  \eqref{Case2_P3_5_const1}, ${\theta_{3n - 2}^{ub}}$ is non-convex w.r.t. ${\bf q}_{3n-2}$, and the constraint set \eqref{Case2_P3_5_const3} is also non-convex. Similarly, by applying the first-order Taylor expansion at  any feasible point ${\bf q}^l_{3n-2}$, we can obtain the lower bound  ${\theta_{3n - 2}^{ub,lb}}$ for ${\theta_{3n - 2}^{ub}}$ in \eqref{Case1_constr1trajectory1NEW} and the similar   result in \eqref{Case1_constr1trajectory2NEW} for the non-convex constraint \eqref{Case2_P3_5_const3}. Thus, we have
\newcounter{mytempeqncnt1}
\begin{figure*}
\normalsize
\setcounter{mytempeqncnt1}{\value{equation}}
\begin{align}
{\log _2}\left( {1 + \frac{{P{a_{3n - 2}}\beta _0^2}}{{\sigma _u^2{{\left( {{t_{3n - 1}}} \right)}^2}}}} \right) \ge {\rm{ }}{\log _2}\left( {1 + \frac{{{C_{3n - 1}}}}{{{{\left( {t_{3n - 1}^l} \right)}^2}}}} \right) - \frac{1}{{\ln 2}}\frac{{2{C_{3n - 1}}}}{{\left( {{{\left( {t_{3n - 1}^l} \right)}^2} + {C_{3n - 1}}} \right){t^l_{3n - 1}}}}\left( {{t_{3n - 1}} - t_{3n - 1}^l} \right)\overset{\triangle}{ =} {\rm{ }}R_{3n - 1}^{u,up}\label{Case2_P3_5_ObjectNEW}
\end{align}
\hrulefill 
\vspace*{4pt} 
\end{figure*}

\begin{align}
&\left( {{\rm{P3}}.6} \right)\mathop {\max }\limits_{{t_{3n - 1}},{s_{3n - 1}},{{\bf{q}}_n}} \sum\limits_{n = 1}^{{N \mathord{\left/
 {\vphantom {N 3}} \right.
 \kern-\nulldelimiterspace} 3}} {{\varphi _{3n - 1}}R_{3n - 1}^{u,up}}  \notag\\
&{\rm s.t.}~\sum\limits_{i = 1}^n {{\varphi _{3i - 1}}\left( {{P_\varepsilon } + \mu {{\log }_2}\left( {1 + \frac{{P{a_{3i - 2}}\beta _0^2}}{{\sigma _u^2{{\left( {{s_{3i - 1}}} \right)}^2}}}} \right)} \right)} \notag\\
 & \qquad\qquad \qquad  \le \sum\limits_{i = 1}^n {\eta \left( {1 - {a_{3i - 2}}} \right)} P\theta_{3i - 2}^{ub,lb},n \in {{\cal N}_2}, \notag\\
 &\qquad{\left\| {{\bf{q}}_{3n - 1}^l - {{\bf{w}}_b}} \right\|^2} + 2{\left( {{\bf{q}}_{3n - 1}^l - {{\bf{w}}_b}} \right)^T} \times \left( {{{\bf{q}}_{3n - 1}} - {\bf{q}}_{3n - 1}^l} \right) \notag\\
 &\qquad\qquad\qquad\qquad\qquad\qquad+ {H^2} \ge {s_{3n - 1}},n \in {\cal N}_2,\notag\\
&\qquad\eqref{Case1_constr2},\eqref{Case2_P3_5_const2}.\notag
\end{align}
The objective function and all the  constraints in problem $(\rm P3.6)$ are convex, and thus it can be efficiently solved by standard convex optimization techniques.  The
details of the overall algorithm for solving $(\rm P3)$ are omitted for brevity, given the similarity to that for $(\rm P1)$.

\begin{figure}[!t]
\centerline{\includegraphics[width=3.5in]{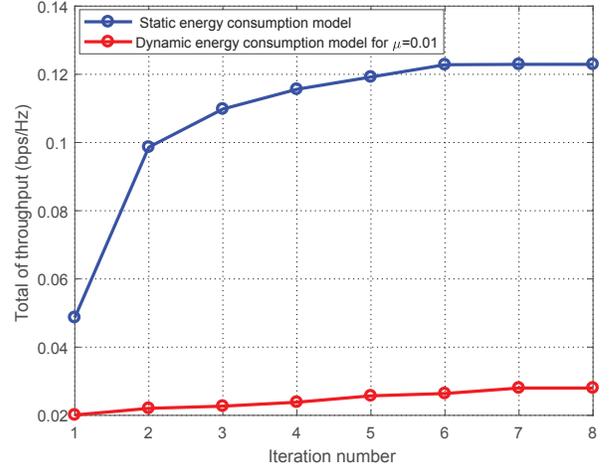}}
\caption{Convergence behaviour of Algorithm 1.} \label{fig2}
\end{figure}

\begin{figure}[!t]
\centerline{\includegraphics[width=3.5in]{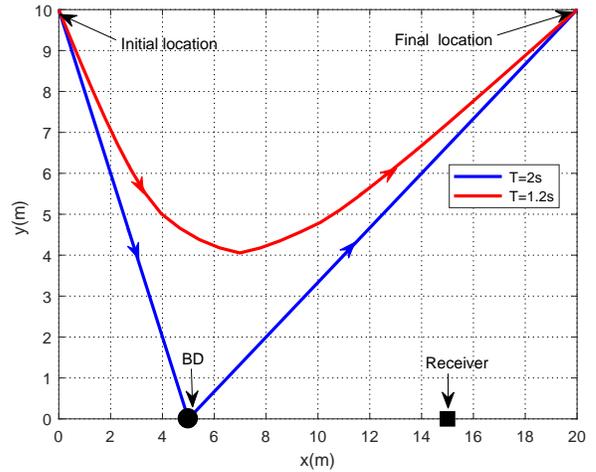}}
\caption{ UAV trajectory obtained by our proposed scheme with  static circuit power consumption cases. } \label{fig3}
\end{figure}

\begin{figure}[!t]
\centerline{\includegraphics[width=3.5in]{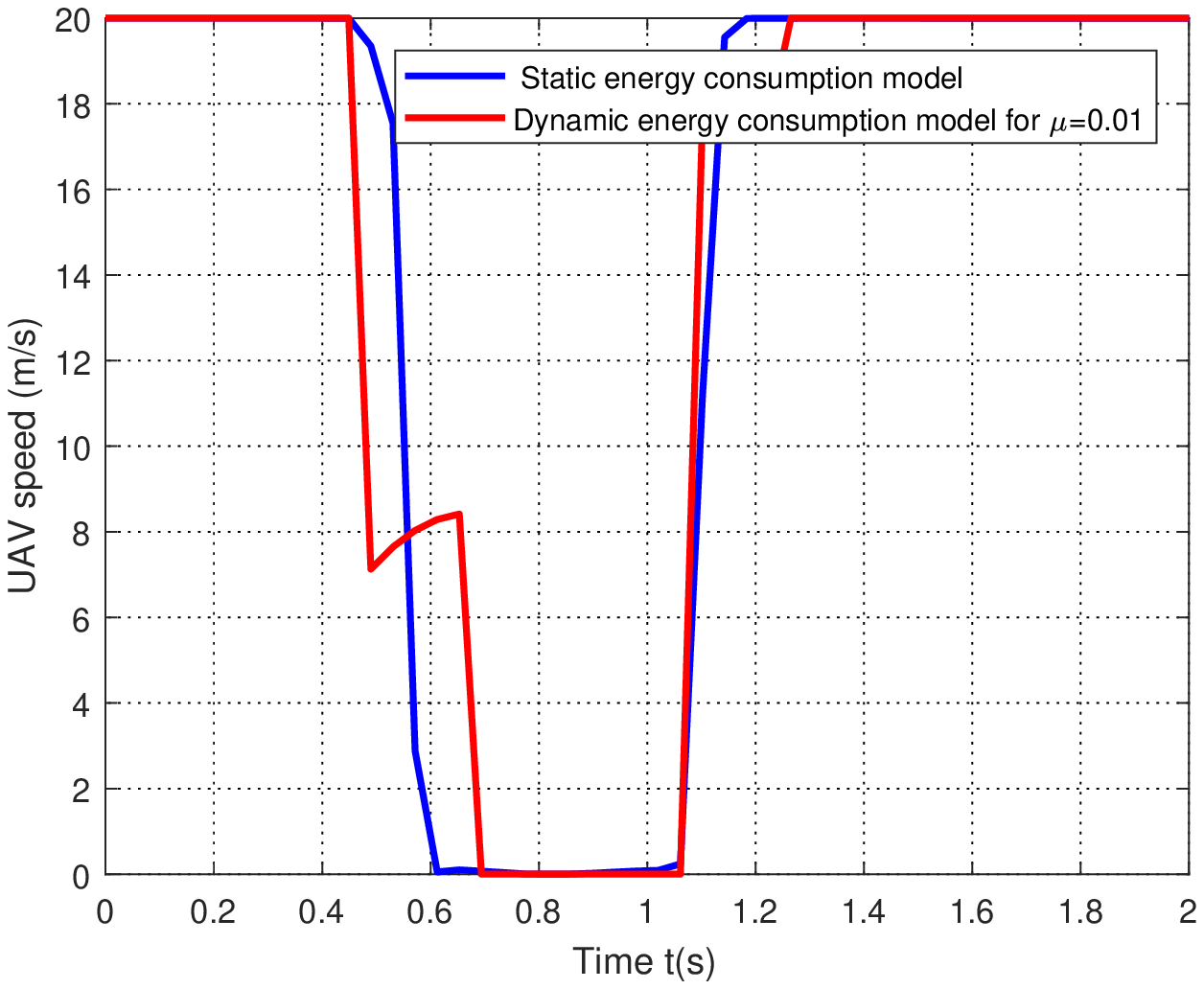}}
\caption{UAV speed  calculated  by our proposed scheme for the two  circuit power consumption cases when  $T=2{\rm s}$.} \label{fig4}
\end{figure}

\begin{figure}[!t]
\centerline{\includegraphics[width=3.5in]{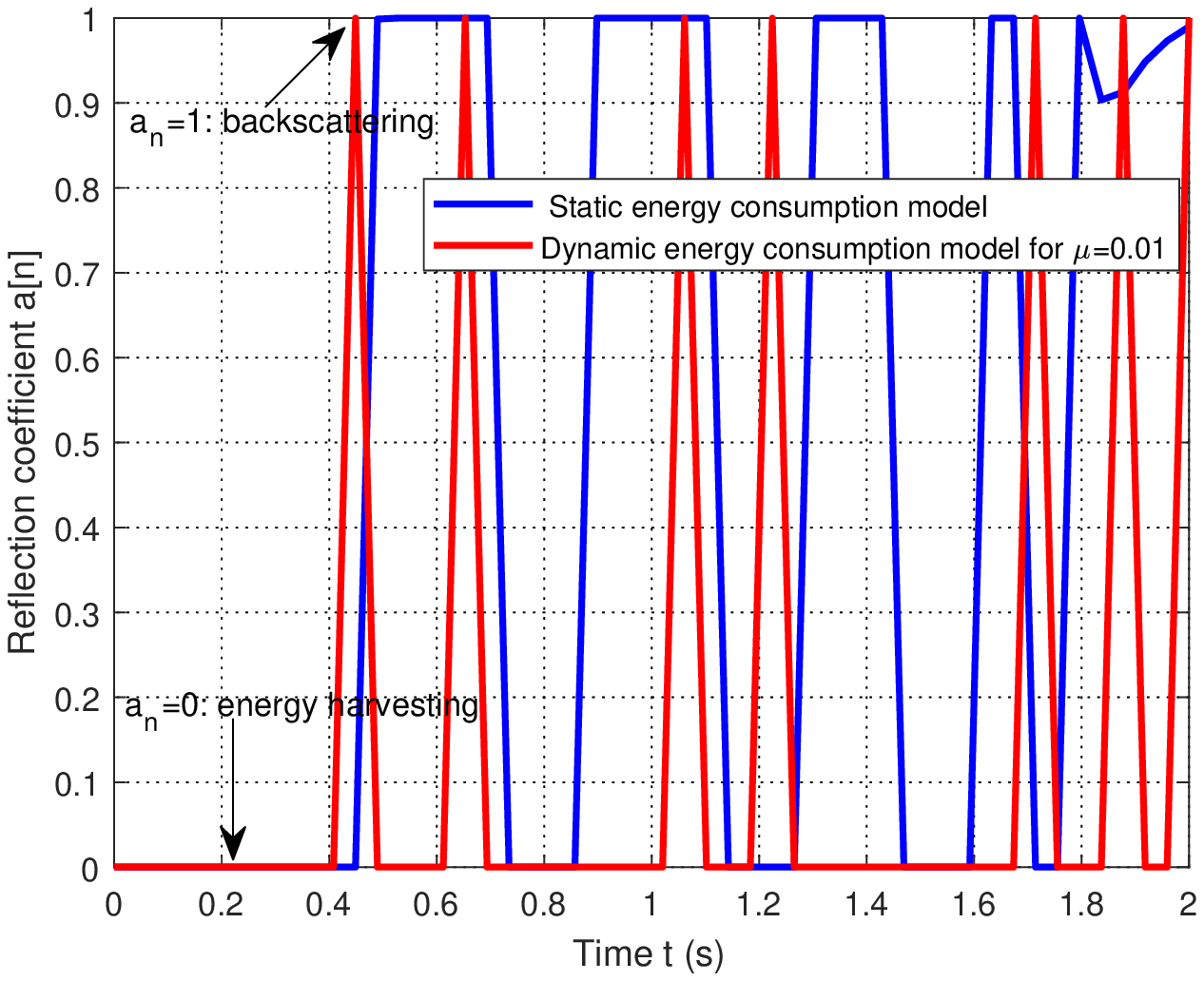}}
\caption{BD reflection coefficient for the two  circuit power consumption cases when  $T=2{\rm s}$.} \label{fig5}
\end{figure}

\begin{figure}[!t]
\centerline{\includegraphics[width=3.5in]{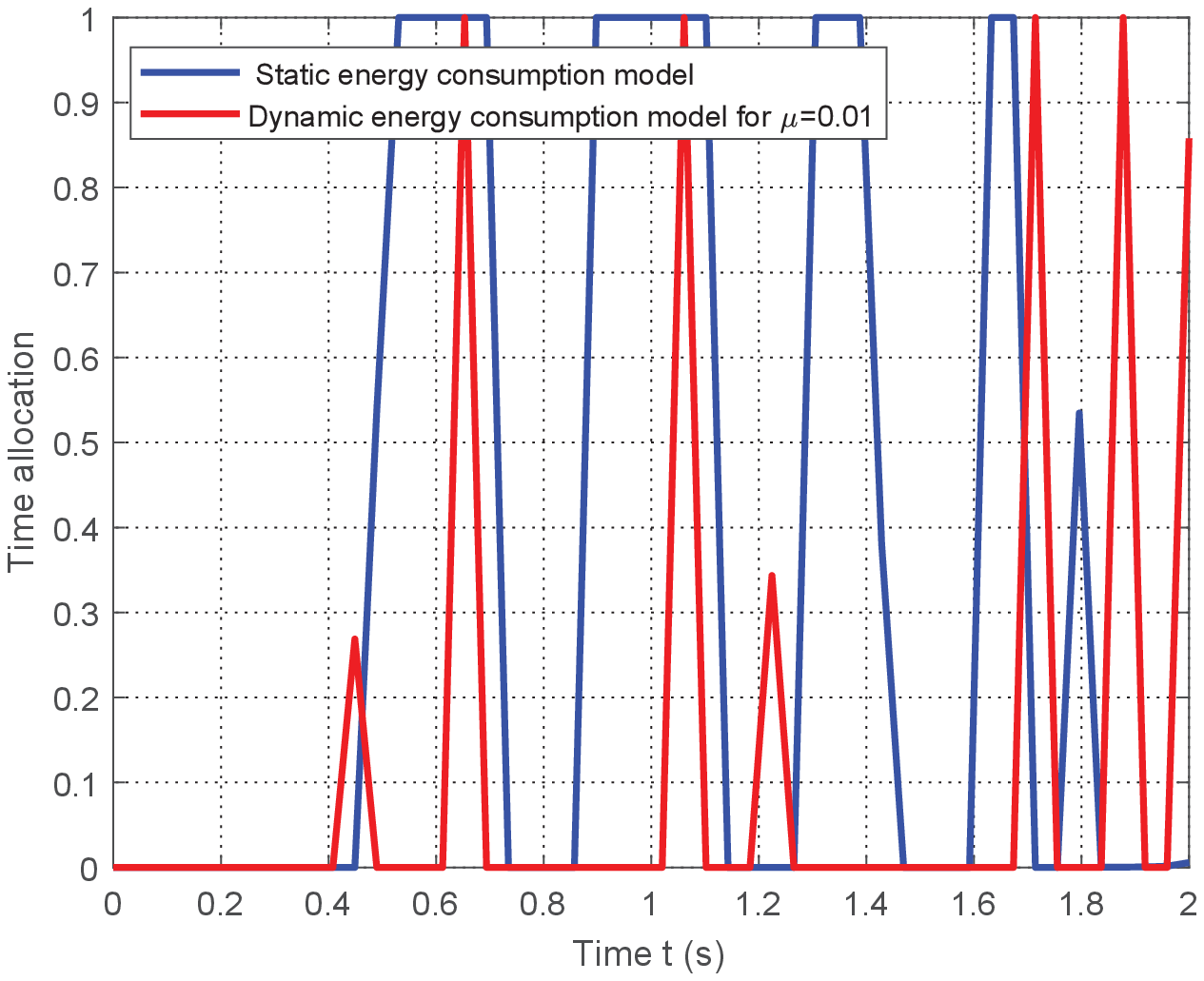}}
\caption{BD transmission time allocation for the  two  circuit power consumption cases when  $T=2{\rm s}$.} \label{fig6}
\end{figure}

\begin{figure}[!t]
\centerline{\includegraphics[width=3.5in]{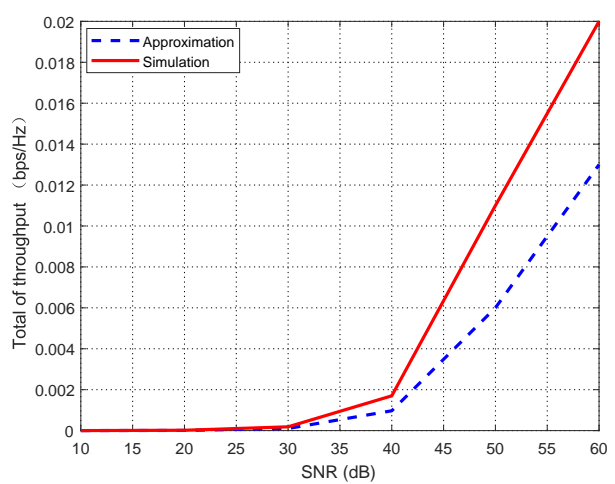}}
\caption{The impact of SNR on system performance based on approximation versus numerical result under  period $T=2\rm s$ and $\mu=0.01$.} \label{fig6_1}
\end{figure}

\begin{figure}[!t]
\centerline{\includegraphics[width=3.5in]{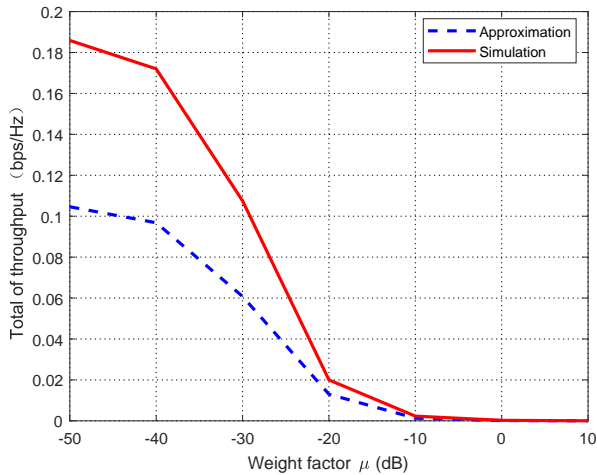}}
\caption{The impact of weight factor $\mu$ on system performance based on approximation versus numerical result under  period $T=2\rm s$.} \label{fig7}
\end{figure}

\begin{figure}[!t]
\centerline{\includegraphics[width=3.5in]{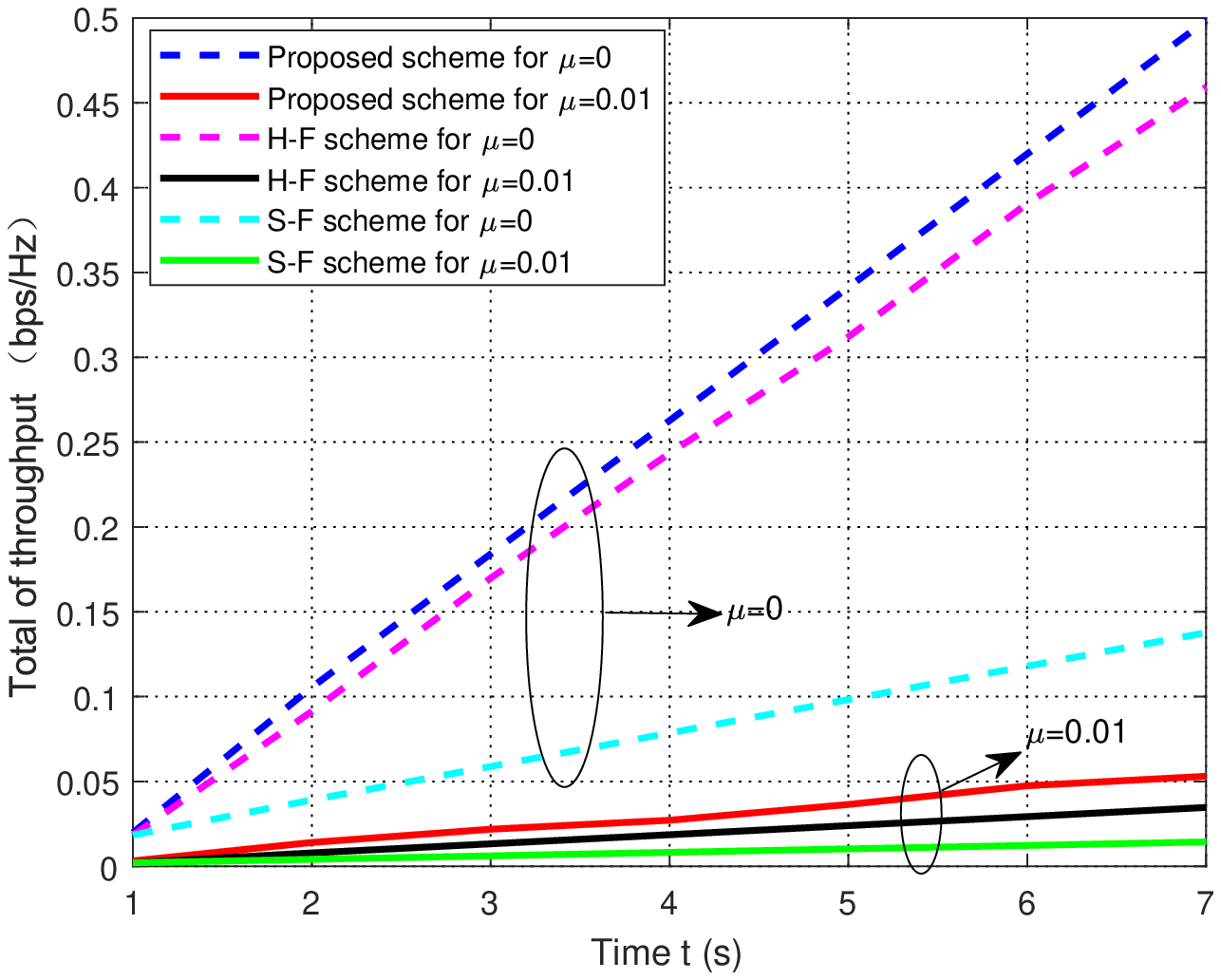}}
\caption{Total of throughput versus period $T$ for  the two  circuit power consumption cases with different optimization schemes.} \label{fig8}
\end{figure}

\section{NUMERICAL RESULTS}
In this section, numerical simulations are provided to evaluate the performance of our proposed schemes. The channel gain of the system is set to ${\beta _0} =  - 30{\rm{dB}}$ \cite{lyu2018relay} \cite{xu2018uav}, and the noise power at the UAV and receiver  are assumed to be equal, $\sigma _r^2 = \sigma _u^2=-90{\rm dB}$ \cite{zhou2018computation}. The UAV altitude is fixed at ${H = 10\rm{m}}$ with the maximum transmit power $P=1{\rm W}$ and maximum speed $V_{\rm max } = 20{\rm{m/s}}$ \cite{lyu2018uav}, \cite{zhou2018computation}. The duration of each time slot is  $\delta=0.04 \rm s$. The energy harvesting efficiency is assumed to be $\eta=0.9$, and the path loss coefficient is set to $m=3$ \cite{lyu2018uav}. The BD's circuit power consumption is in the order of micro-watt \cite{liu2013ambient}, \cite{iyer2016inter}, we set $P_c=10^{-5} \rm W$ and ${P_\varepsilon } = \frac{{{P_c}}}{{5}}$ without loss of generality. The step size $\pi$ is set to 0.01. The horizontal locations of the  BD and receiver are respectively set to ${\bf w}_b=(5{\rm m},0)^T$ and ${\bf w}_r=(15{\rm m},0)^T$. The UAV's  initial and final location are ${\bf q}_{\rm I}=\left( {0,10{\rm m}} \right)^T$ and ${\bf q}_{\rm F}=\left( {20{\rm m},10{\rm m}} \right)^T$, respectively.
\subsection{Direct Link Available}
We first consider the system model where the direct link between BD and receiver is available. In our initial setups, the initial trajectory for UAV is a straight path from the initial location to the final location with a steady speed, and the initial reflection coefficient for BD at any time slot $n$ is set to 0.5. In Fig.~\ref{fig2}, we plot  the convergence behaviour of Algorithm~\ref{alg1} for the static and dynamic circuit power consumption when $T=2\rm s$. It is observed that  the  throughput increases quickly  and converges within a few iterations, which demonstrates the effectiveness of the Algorithm~\ref{alg1}.

Fig.~\ref{fig3} shows the UAV trajectory obtained for the static circuit power consumption model for $T=2\rm s$ (the trajectory for the dynamic  model is essentially identical and thus is omitted). We see that the UAV flies in a straight line to the BD, and then directly to the final location. The corresponding UAV speed  for both circuit power  models are plotted in Fig.~\ref{fig4}. For the static model, we see that the UAV first flies with maximum speed towards the  BD,   hovers above the BD, and then flies with maximum speed from the BD to the final location.  The UAV moves in the direction of the BD to improve the throughput, and has time to hover hear the BD before moving towards the final location. The maximum allowed hover time is constrained by the maximum UAV speed, the distance from the initial/final location  to the BD, and the period $T$.

The optimized reflection coefficients for the two circuit power models are plotted in Fig.~\ref{fig5} for the case of  for period $T=2 \rm s$. A value of $a_n=0$ indicates that the BD only  harvests energy from the UAV at time slot $n$, and  no data is backscattered by the  BD. In contrast, $a_n=1$ indicates that  the BD transmits  the data to receiver with maximum reflection coefficient , and no energy is harvested.  It is observed that for the static circuit power  model, the BD harvests energy first, and then  uses the preserved energy for data transmission with maximum reflection coefficient. Interestingly,  the duration for  BD data backscattering for the static circuit power consumption model is significantly larger than the  dynamic model. This is due to the fact that the dynamic  model is more energy hungry than the static  model, and  as a consequence, more time needed to  harvest  energy from the UAV. In addition, the optimized  time allocation for the  BD is  plotted in Fig.~\ref{fig6}. When  $a_n=1$ within period from $0.4\rm s$ to $1.7\rm s$ in Fig.~\ref{fig5}, the entire time slot $n$ will be used for data backscattering for the static circuit power  model  as shown in Fig.~\ref{fig6}. In contrast, for  $a_n=0$,   time slot $n$ will not be used for  backscattering. Also, for the dynamic circuit power  model, the BD's backscattering time is much less than the static circuit power  model case.

To evaluate  the accuracy of the approximation of the expected throughput given in \eqref{Case1_Ratelower},  the desired throughput given in \eqref{Case1_expression_2} is compared.  Fig.~\ref{fig6_1} shows the compared results  for the different SNR under period $T=2 \rm s$ and $\mu=0.01$. The value SNR is calculated as $\beta_0/\sigma_r^2$ with fixed ${\beta _0} =  - 30{\rm{dB}}$ and $\sigma _r^2$ varying from $-40{\rm dB}$ to $-90{\rm dB}$.  For the approximation of the expected throughput, the results are obtained via Algorithm \ref{alg1}. For the desired throughput,  the desired throughput is averaged over $10^6$ random channel realizations at each UAV location under $K=15\rm{dB}$. It is expected that  the throughput is monotonically increasing with SNR for the  approximation and simulation realizations. In addition, it also can be seen that   for the  small value  SNR, namely $40{\rm{dB}}$, the obtained throughput for approximation is almost same as for numerical simulation. For the relatively large value SNR, the proposed optimization technique based on approximation may still be applied with an acceptable  accuracy.

Fig.~\ref{fig7} shows the results  of   approximation and numerical simulations for the different weight factor $\mu$ under period $T=2 \rm s$. First, it is observed that the system throughput degrades as $\mu$ increases for the approximation and numerical simulations. This is expected since a larger $\mu$ means more dynamical energy must be  consumed, and hence the time allocation and reflection coefficient need to be smaller to satisfy the energy harvesting and circuit power consumption  constraint. Second, for  a very small weight factor with $\mu=-50\rm{dB}$,  the approximation of the expected throughput of $0.11 {\rm {bps/Hz} }$  is achieved. When the weight factor $\mu$ is higher than $-20\rm{dB}$, the approximation achieves a good accuracy with the  numerical simulations.

In Fig.~\ref{fig8}, to show the superiority of our proposed scheme, we compare it with the  following  benchmarks: 1) H\text{-}F scheme, where the BD  forwards the data at time slot $n$ by using the harvested energy from time slot $n-1$. Mathematically, the  constraint \eqref{Case1_constr1} is changed  to ${\varphi _{2i}}P_{2i}^e \le \eta \left( {1 - {a_{2i - 1}}} \right)P\theta_{2i - 1}^{ub}, n\in {\cal N}_1 $; 2) S\text{-}F scheme, where the UAV flies directly from the  initial location  to the final location  in a straight line, but the time allocation and reflection coefficient are both optimized.   It is observed that the  static  model results in a higher performance gain compared with the dynamic model for three schemes. This is expected since the higher transmission means more  energy will be consumed, which thus degrades the system performance due to the energy budget constraint. In addition, the proposed scheme and the H\text{-}F scheme outperform  the S\text{-}F scheme, which shows the advantage of optimizing the UAV trajectory in order to realize the full benefit of the UAV-aided backscatter communication.
\begin{figure}[!t]
\centerline{\includegraphics[width=3.5in]{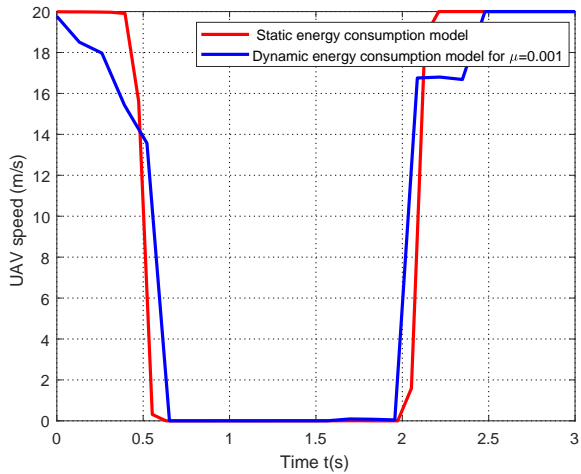}}
\caption{UAV speed  obtained by our proposed scheme for the two  circuit power consumption cases when $T=3{\rm s}$.} \label{fig9}
\end{figure}

\begin{figure}[!t]
\centerline{\includegraphics[width=3.5in]{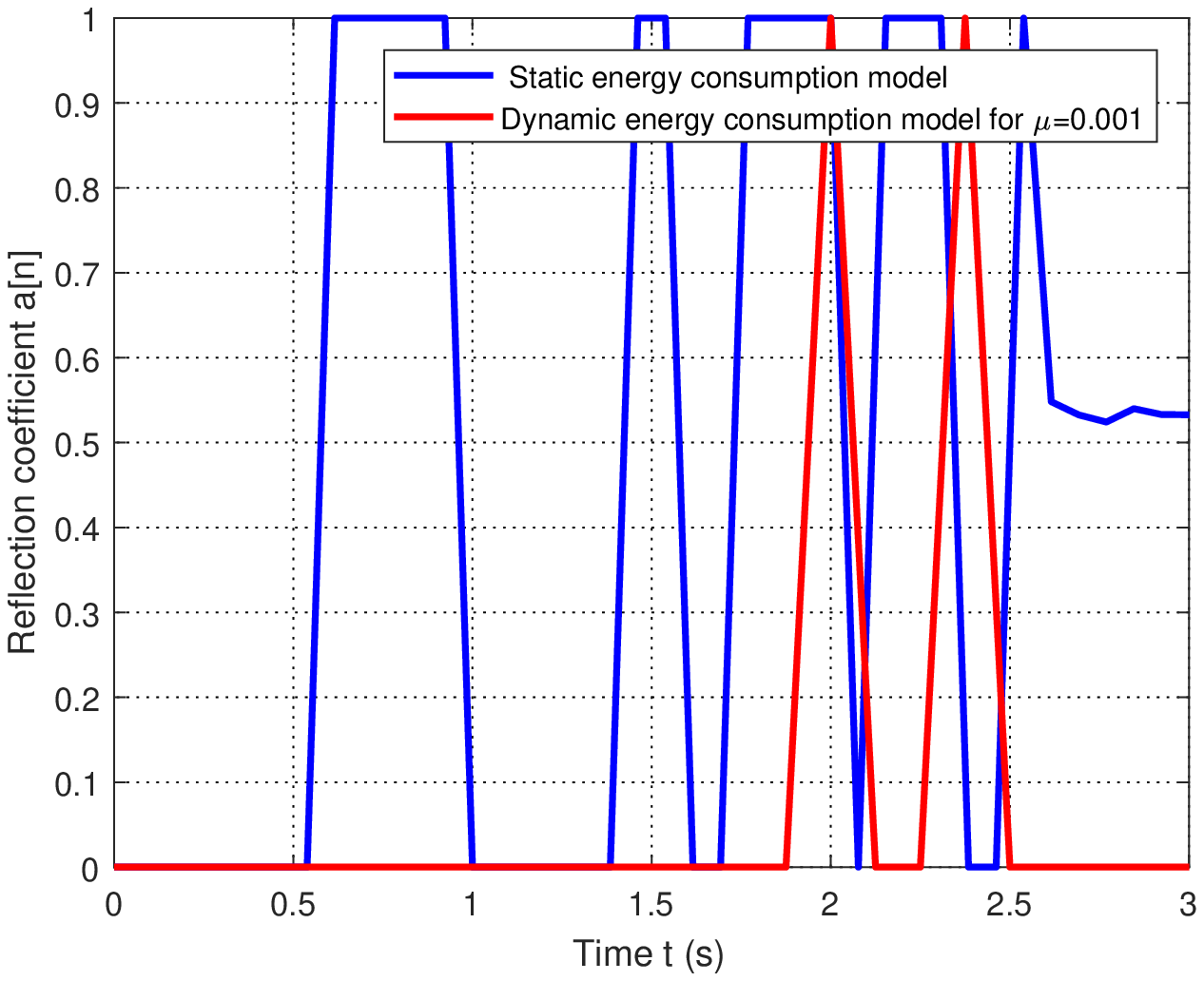}}
\caption{BD reflection coefficient  for the two  circuit power consumption cases when $T=3{\rm s}$.} \label{fig10}
\end{figure}

\begin{figure}[!t]
\centerline{\includegraphics[width=3.5in]{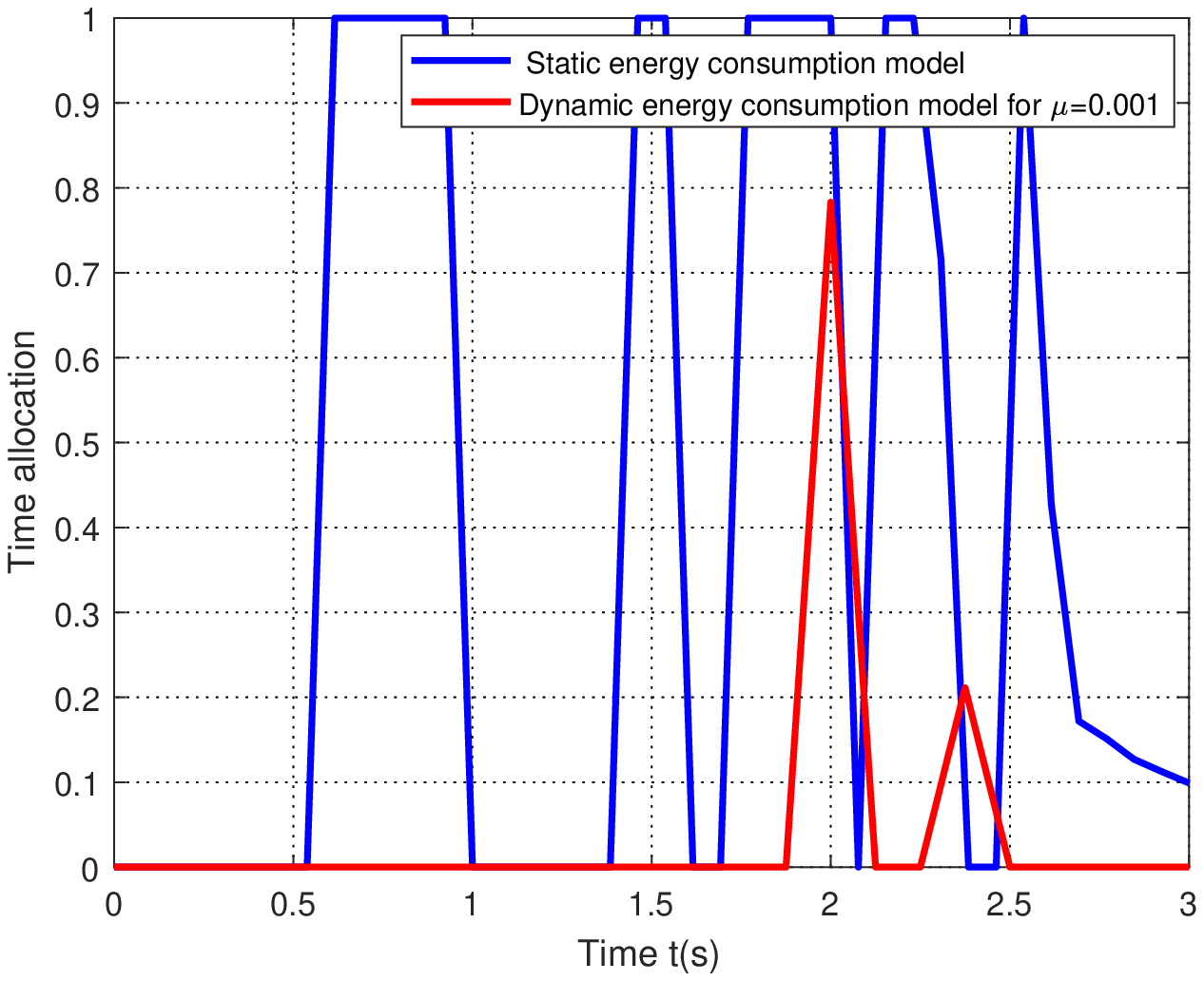}}
\caption{BD  transmission time allocation  for the two  circuit power consumption cases when $T=3{\rm s}$.} \label{fig11}
\end{figure}

%

\begin{figure}[!t]
\centerline{\includegraphics[width=3.5in]{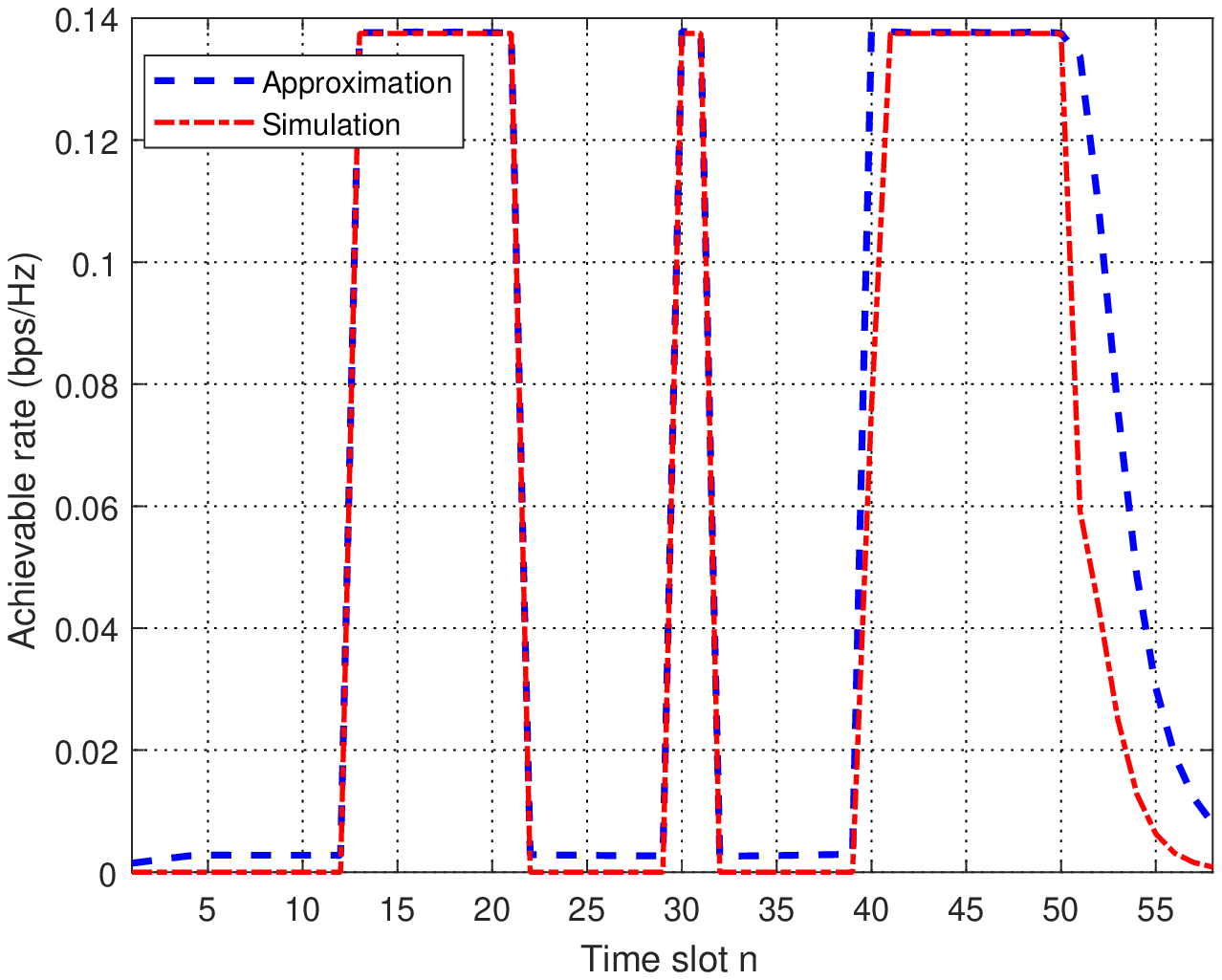}}
\caption{Approximation result based on \eqref{Case2_RU_approxi} versus numerical simulation based on \eqref{Case2_RU}  under $\mu=0$, $K=15\rm dB$, and  $T=7\rm s$.} \label{fig12_1}
\end{figure}

\begin{figure}[!t]
\centerline{\includegraphics[width=3.5in]{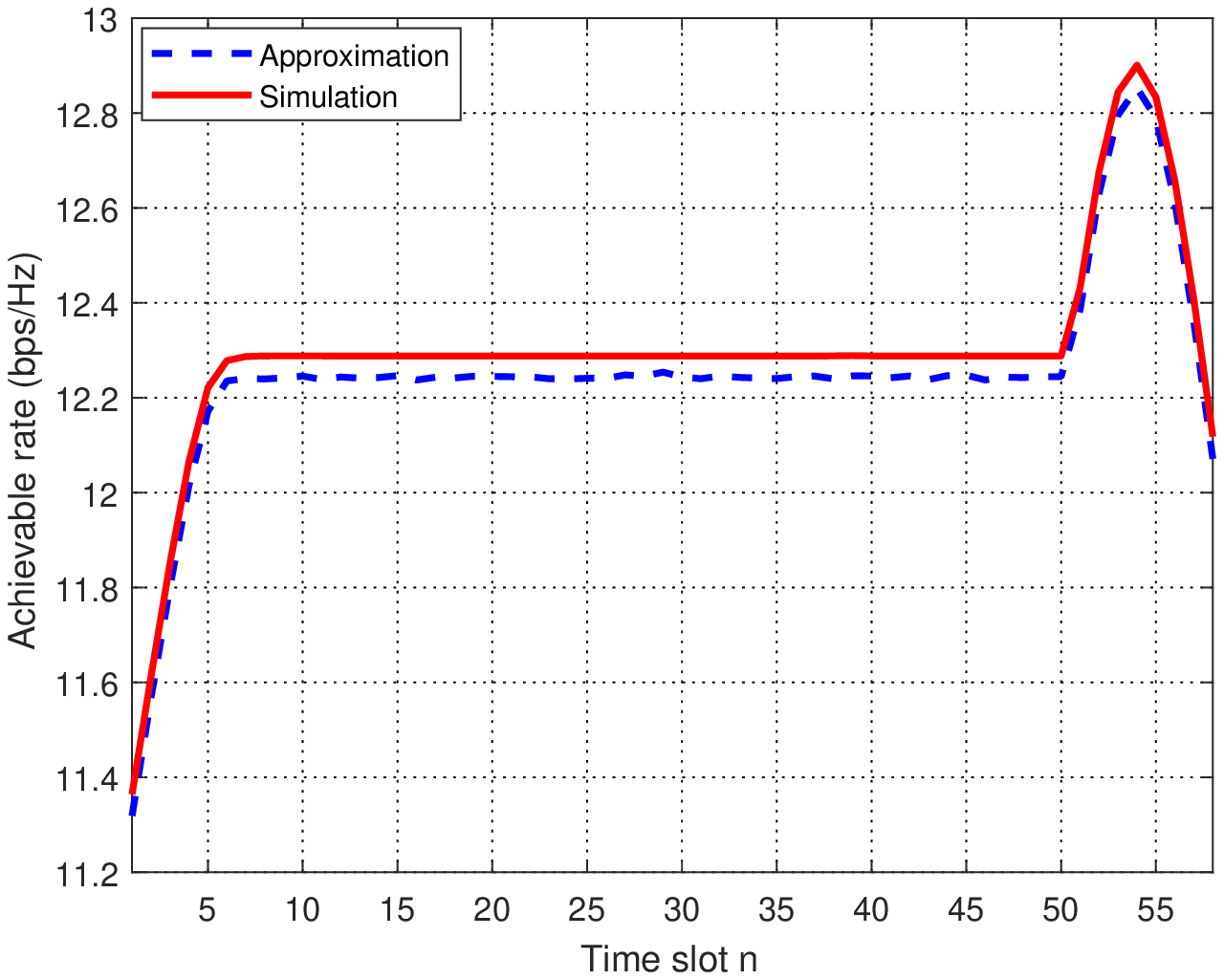}}
\caption{Approximation based on \eqref{Case2_Rr_approxi} versus numerical simulation based on \eqref{Case2_Rr_simulation} under $\mu=0$, $K=15\rm dB$, and  $T=7\rm s$.} \label{fig12_2}
\end{figure}

\begin{figure}[!t]
\centerline{\includegraphics[width=3.5in]{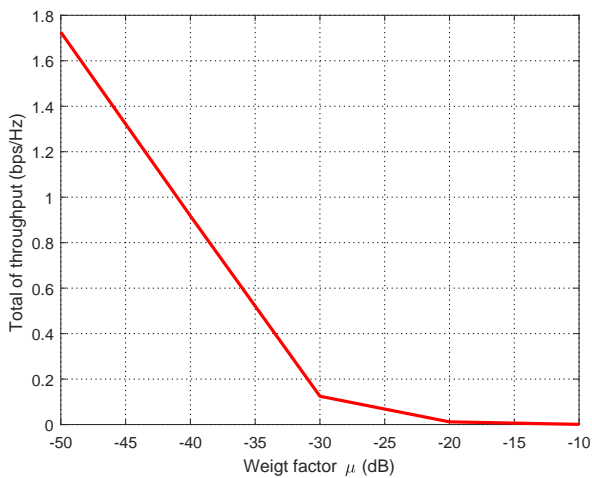}}
\caption{The impact of the  weight  factor $\mu$ on system performance for period $T=3\rm s$.} \label{fig12}
\end{figure}

\begin{figure}[!t]
\centerline{\includegraphics[width=3.5in]{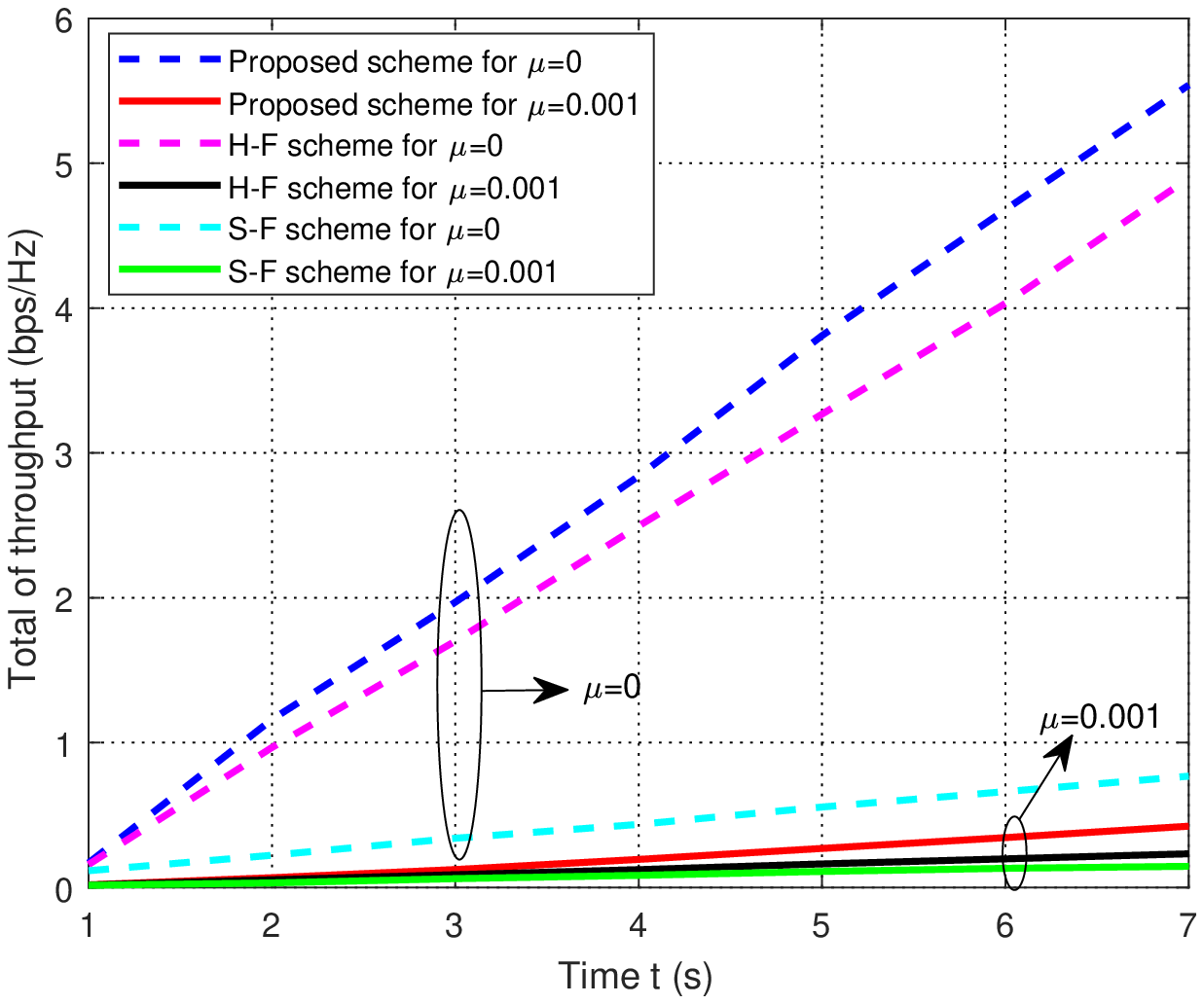}}
\caption{Total of throughput versus period $T$ for two  circuit power consumption cases with different optimization schemes.} \label{fig13}
\end{figure}

\subsection{Direct Link Unavailable}
Next, we study the throughput maximization of the backscatter network when the direct link between BD and receiver is not available. Fig.~\ref{fig9} shows the UAV speed obtained by our proposed scheme  for both the static and dynamic power models when  $T=3\rm s$. The UAV trajectories for both models are  nearly the same as  the UAV trajectory plotted in Fig.~\ref{fig3}. For the static power model in Fig.~\ref{fig9}, the behavior of the UAV is the same as before, flying with maximum speed to the BD, hovering above the BD for as long as possible, then flying with maximum speed to the final location. This is expected since the longer the BD can be served, more energy can be harvested and the system throughput will increase. The result for the dynamic model is similar.

Figs.~\ref{fig10} and \ref{fig11} show the   reflection coefficient and time allocation obtained by our proposed schemes, respectively. For both  circuit power models, the BD first harvests energy to the UAV and then reflects signal to the UAV. We can also see that the energy harvesting time for the dynamic circuit power model is  larger than in the static model, which indicates that the dynamic model results in less time for  BD data backscattering.

Fig.~\ref{fig12_1} and  Fig.~\ref{fig12_2} compare  the approximation of expected achievable rate  and the numerical simulation of  expected achievable rate under  $\mu=0$, $K=15\rm dB$, and  $T=7\rm s$. Fig.~\ref{fig12_1} first shows the results of achievable rate  for approximation based on $R^u_{3n-1}$ in \eqref{Case2_RU_approxi} and numerical simulation based on $\check R^u_{3n-1}$ in \eqref{Case2_RU}.   It is observed from Fig.~\ref{fig12_1} that the approximation result $R^u_{3n-1}$ matches well with the  numerical simulation result $\check R^u_{3n-1}$ at  all time slot. Fig.~\ref{fig12_2} shows the results of achievable rate  for approximation based on  $R_{3n}^r$ in \eqref{Case2_Rr_approxi} and numerical simulation based on $\check R_{3n}^r$ in \eqref{Case2_Rr_simulation}. It is observed from Fig.~\ref{fig12_2} that  a satisfactory accuracy for approximation result $R_{3n}^r$ and  numerical simulation result $\check R_{3n}^r$ is also obtained. In addition, comparing with Fig.~\ref{fig12_1} and Fig.~\ref{fig12_2}, the value $R^u_{3n-1}$ shown in Fig.~\ref{fig12_1}  is indeed much smaller than  $R^r_{3n}$ shown in Fig.~\ref{fig12_2}. This demonstrates the valid  assumption of Remark~\ref{remark1}. In addition, for the non-zero weight factor $\mu$, we can obtain similar result as in the case of $\mu=0$, and is omitted here for brevity.

Fig.~\ref{fig12} shows the  impact of the weight factor $\mu$ on the system performance. The throughput is monotonically decreasing with the value $\mu$. For example, with a small value $\mu=-50 \rm dB$, the throughput can achieve up to $1.76\rm bps/Hz$. However, when $\mu$ is larger than $\mu=-20 \rm dB$, the system throughput is nearly to zero.

In Fig.~\ref{fig13}, we compare the system throughput achieved by our proposed scheme with the other benchmarks for different values of  $T$.  We see that for $\mu=0$, our proposed scheme is superior to the  H\text{-}F scheme and S\text{-}F scheme achieving higher throughput than the  benchmarks. The performance gain becomes more substantial as $T$ grows. In addition, both the proposed scheme and H\text{-}F scheme outperform the S\text{-}F  scheme, which indicates that optimization of the UAV trajectory significantly improves the system performance. Furthermore, it is observed that the schemes with  static circuit power model for $\mu=0$ obtain a larger system throughput than the dynamic  model for  $\mu=0.001$.

\section{Conclusion}
This paper studied a UAV-aided BackCom network   with and without a direct link between BD and receiver. Different static/dynamic circuit power  consumption models for the two system models were  considered. By exploiting the UAV mobility, the end-to-end achievable rate was maximized by jointly optimizing the time allocation, reflection coefficient and UAV trajectory. By means of the block coordinate descent and SCA techniques, an efficient iterative algorithm was proposed for both system models.  The optimal   time allocation  for a  given UAV trajectory under the static  circuit power consumption model  was derived in closed-form. In addition, for this case the optimal  reflection coefficient was obtained with  low computational complexity by using the Lagrangian dual method.  Simulation  results showed that the UAV mobility is beneficial for achieving a much higher system  throughput  than the other benchmarks that do not consider trajectory optimization. In addition, it was shown that more time will be used to backscatter for static circuit power consumption model compared with the dynamic  circuit power consumption model, and  results in  much higher throughput  than the dynamic  model. Finally, it was shown that the proposed scheme significantly  outperforms the H-F based  scheme, thanks to the more degree of  freedom for performance enhancement  via careful   energy harvesting and circuit power consumption design. The results in this paper can be further extended by considering following research directions: 1) The  limited buffer size on BD; 2) Joint UAV trajectory and  multiple access design for the  multiple backscatter devices scenario; 3) The study of   energy-efficient fixed/rotary  wing UAV trajectory design  by taking into account the UAV propulsion energy consumption.

\appendices
\section{Proof of Theorem~\ref{theorem1} } \label{appendix1}
To show Theorem~\ref{theorem1}, we first define the function $f\left( x \right){\rm{ = }}{{\mathbb{E}}_{{X_1}{X_2}}}\left[ {{{\log }_2}\left( {1 + {x_1}{x_2}} \right)} \right],({x_1},{x_2} > 0)$, where $x_1$ and $x_2$ are independent with each other. We then have
\begin{align}
f\left( x \right)\overset{a}{\le} {\mathbb E_{{X_1}}}\left\{ {\left[ {{{\log }_2}\left( {1 + {x_1}{\mathbb E_{{X_2}}}\left[ {{x_2}} \right]} \right)} \right]} \right\} = \mathord{\buildrel{\lower3pt\hbox{$\scriptscriptstyle\frown$}}
\over f} \left( x \right),\label{append0}
\end{align}
where inequality (a) in \eqref{append0} holds due to the concavity of ${{{\log }_2}\left( {1 + {x_1}{x_2}} \right)}$ w.r.t. $x_2$ and   Jensen's inequality.

Based on the convexity of ${{{\log }_2}\left( {1 + e^x} \right)}$ w.r.t. $x$ and   Jensen's inequality, we  have
\begin{align}
\mathord{\buildrel{\lower3pt\hbox{$\scriptscriptstyle\frown$}}
\over f} \left( x \right) &= {\mathbb E_{{X_1}}}\left[ {{{\log }_2}\left( {1 + {x_1}{\mathbb E_{{X_2}}}\left[ {{x_2}} \right]} \right)} \right] \notag\\
&= {\mathbb E_{{X_1}}}\left[ {{{\log }_2}\left( {1 + {\mathbb E_{{X_2}}}\left[ {{x_2}} \right]{e^{\ln {x_1}}}} \right)} \right]\notag\\
&\ge {\log _2}\left( {1 + {\mathbb E_{{X_2}}}\left[ {{x_2}} \right]{e^{{\mathbb E_{{X_1}}}\left[ {\ln {x_1}} \right]}}} \right) = \hat f\left( x \right).
\end{align}
We should point out that $\hat f\left( x \right)$ is neither  a upper bound result nor a lower bound result for $\mathord{\buildrel{\lower3pt\hbox{$\scriptscriptstyle\frown$}}
\over f} \left( x \right)$. Instead,  $\hat f\left( x \right)$ is served as an approximation result for $\mathord{\buildrel{\lower3pt\hbox{$\scriptscriptstyle\frown$}}
\over f} \left( x \right)$.  Letting  ${x_1} = \frac{{P{a_n}{h_{br}}}}{{\sigma _r^2}}$ and ${x_2} = h_n^{ub}$, we have
\begin{align}
{\mathbb E_{{X_2}}}\left[ {{x_2}} \right]{\rm{ = }}\mathbb E\left[ {\theta _n^{ub}{{\| {\tilde h_n^{ub}} \|}^2}} \right]{\rm{ = }}\frac{{{\beta _0}}}{{{{\left\| {{{\bf{q}}_n} - {{\bf{w}}_f}} \right\|}^2} + {H^2}}}\label{append2}
\end{align}
and
\begin{align}
{\mathbb E_{{X_1}}}\left[ {\ln {x_1}} \right] = \ln \frac{{P{a_n}{\beta _0}d_{br}^{ - m}}}{{\sigma _r^2}} +\mathbb E\left[ {\ln \xi } \right] = \ln \frac{{P{a_n}{\beta _0}d_{br}^{ - m}}}{{\sigma _r^2}} - {\kappa _0}, \label{append2.1}
\end{align}
where ${\kappa _0}$ is the Euler constant, $\mathbb E\left[ {\ln \xi } \right]$ is derived from eq.(4.331.1) in \cite{gradshteyn2014table}.
Substituting \eqref{append2} and \eqref{append2.1} into $\hat f\left( x \right)$, we can easily obtain $\hat R^{r}_ {n + 1}$ in \eqref{Case1_Ratelower}. This completes the  proof of Theorem~\ref{theorem1}.

\section{Proof of Theorem ~\ref{theorem2} } \label{appendix2}
We first observe that problem $(\rm P1.1)$ is a linear optimization problem. Meanwhile, it can be verified that $(\rm P1.1)$  satisfies  Slater's condition, and thus the dual gap is zero and the optimal solution can be obtained by solving its dual problem \cite{boyd2004convex}. Let  ${\lambda _n}>0$ for $n=1,...,N/2$ be the Lagrangian dual variables  corresponding to \eqref{Case1_constr1}. The corresponding partial Lagrangian for problem $(\rm P1)$ can be expressed as
\begin{align}
&{\cal L}\left( {{\varphi _{2n}},{\lambda _n}} \right) = \sum\limits_{n = 1}^{N/2} {{\varphi _{2n}}\hat R_{2n}^{r}}  + \notag\\
&\sum\limits_{n = 1}^{N/2} {{\lambda _n}\left( {\sum\limits_{i = 1}^n {\eta \left( {1 - {a_{2i - 1}}} \right)} P\theta_{2i - 1}^{ub} - \sum\limits_{i = 1}^n {{\varphi _{2i}}{P_c}} } \right)} . \label{append3}
\end{align}
The KKT conditions are sufficient to obtain the optimal solution, and the partial conditions are given by
\begin{align}
&\frac{{\partial {\cal L}\left( {\varphi _{2n}^*,\lambda _n^*} \right)}}{{\partial \varphi _{2n}^*}} = 0, \label{append4}\\
&\lambda _n^*\left( {\sum\limits_{i = 1}^n {\eta \left( {1 - {a_{2i - 1}}} \right)} P\theta_{2i - 1}^{ub} - \sum\limits_{i = 1}^n {\varphi _{2i}^*{P_c}} } \right) = 0.\label{append5}
\end{align}
From equation \eqref{append4}, the optimal dual variables can be obtained  as
\begin{align}
&\lambda _n^* = \frac{{\hat R_{2n}^{r} - \hat R_{2n + 2}^{r}}}{{{P_c}}},n = 1,...,N/2 - 1\notag\\
&\lambda _{N/2}^* = \frac{{\hat R_N^{r}}}{{{P_c}}}.
\end{align}
If  ${\hat R_{2n}^{r}}$ is a decreasing function with  $n\in {\cal N}_1$, the optimal value is  positive, i.e., $\lambda _n^*>0$ for $n=1,...,N/2$. Based on the complementary slackness condition from \eqref{append5}, we must have
\begin{align}
\sum\limits_{i = 1}^n {\eta \left( {1 - {a_{2i - 1}}} \right)} P\theta_{2i - 1}^{ub} - \sum\limits_{i = 1}^n {\varphi _{2i}^*{P_c}}  = 0,n \in {{\cal N}_1}.\label{append6}
\end{align}
Then, with  \eqref{Case1_constr4} and  \eqref{append6}, the optimal time allocation $\varphi_{2n}$ can be readily obtained in \eqref{Case_1timeallocation}. This  completes the  proof of Theorem~\ref{theorem2}
\section{Proof of Lemma ~\ref{lemma1} } \label{appendix3}
Similar to Appendix~\ref{appendix2}, let  ${\nu _n}>0$ for $n=1,...,N/2$ be the Lagrangian dual variables  corresponding to \eqref{Case1_constr1}. The corresponding partial Lagrangian for problem $(\rm P1.2)$ can be expressed as
\begin{align}
&{\cal L}\left( {{a_{2n - 1}},{\nu _n}} \right) = \sum\limits_{n = 1}^{N/2} {{\varphi _{2n}}\hat R_{2n}^{r}}   + \notag\\
&\sum\limits_{n = 1}^{N/2} {{\nu _n}\left( {\sum\limits_{i = 1}^n {\eta \left( {1 - {a_{2i - 1}}} \right)} P\theta_{2i - 1}^{ub} - \sum\limits_{i = 1}^n {{\varphi _{2i}}{P_c}} } \right)} . \label{append7}
\end{align}
Accordingly, the dual function for $(\rm P1)$ is given by
\begin{align}
g\left( {{v_n}} \right) = \left\{ \begin{array}{l}
\mathop {\max }\limits_{a_{2n-1}} {\cal L}\left( {a_{ 2n-1},{v_n}} \right)\\
\quad{\rm s.t.}~ \eqref{Case1_constr3}.
\end{array} \right.\label{append8}
\end{align}
By applying the first order derivative  of  \eqref{append7} w.r.t $a_{2n-1}$ and setting it to zero,, we have
\begin{align}
&\frac{{{W_{br}}{\varphi _{2n}}}}{{\ln 2\left( {{{\left\| {{{\bf{q}}_{2n - 1}} - {{\bf{w}}_b}} \right\|}^2} + {H^2}} \right) + {W_{br}}{a_{2n - 1}}}} \notag\\
&\qquad\qquad\qquad\qquad\qquad\qquad-\eta P\theta_{2n - 1}^{ub}\sum\limits_{i = n}^{N/2} {{\nu _i}}  = 0. \label{append9}
\end{align}
Then, the formula \eqref{Case_1statictimeallocation} can be readily obtained based on \eqref{Case1_constr3} and \eqref{append9}. This  completes the  proof of Lemma~\ref{lemma1}.

\bibliographystyle{IEEEtran}
\bibliography{Tcom}

\end{document}